\documentclass[11pt,a4paper]{article}
\usepackage{graphicx}
\usepackage{epstopdf}
\usepackage{amstext,amsfonts,amsbsy,amssymb,bbm}
\usepackage{pdfsync}
\usepackage{a4wide}
\usepackage{times}

\newcommand{\Img}{\mathop{\mbox{Img}}}
\newcommand{\Ker}{\mathop{\mbox{Ker}}}

\def\Tr{\mathop{\rm Tr}\nolimits}

\newcommand {\be}[1]{\begin{eqnarray} \mbox{$\label{#1}$}  }
\newcommand{\ee}{\end{eqnarray}}

\newcommand{\iden}{\mathbbm{1}}

\newcommand{\cD}{ {\cal D} }

\newcommand{\cP}{ {\cal P} }

\newcommand{\cS}{ {\cal S} }

\newcommand{\ga}{ {\alpha} }
\newcommand{\gb}{ {\beta} }

\newcommand{\gd}{ {\delta} }

\newcommand{\gs}{ {\sigma} }

\renewcommand{\ge}{ {\epsilon} }
\newcommand{\gl}{ {\lambda} }
\newcommand{\gr}{\rho}
\newcommand{\gt}{\tau}

\newcommand{\wt}[1]{\widetilde{#1}}

\begin{document}

 \bibliographystyle{unsrt}




\title{Low rank positive partial transpose states and their relation
to product vectors}

\author{Leif Ove Hansen$^a$, Andreas Hauge$^a$, Jan Myrheim$^a$,
and Per {\O}yvind Sollid$^b$ \\
${(a)}$ Department of Physics,
Norwegian University of Science and Technology,\\
N-7491 Trondheim, Norway\\
${(b)}$ Department of Physics, University of Oslo,\\
N-0316 Oslo, Norway}



\maketitle

\begin{abstract}
  It is known that entangled mixed states that are positive under
  partial transposition (PPT states) must have rank at least four.  In
  a previous paper we presented a classification of rank four
  entangled PPT states which we believe to be complete.  In the
  present paper we continue our investigations of the low rank
  entangled PPT states.  We use perturbation theory in order to
  construct rank five entangled PPT states close to the known rank
  four states, and in order to compute dimensions and study the
  geometry of surfaces of low rank PPT states.  We exploit the close
  connection between low rank PPT states and product vectors.  In
  particular, we show how to reconstruct a PPT state from a sufficient
  number of product vectors in its kernel.  It may seem surprising
  that the number of product vectors needed may be smaller than the
  dimension of the kernel.
\end{abstract}


\section{Introduction}

Quantum entanglement between subsystems of a composite physical system
is a phenomenon which clearly distinguishes quantum physics from
classical physics \cite{RPMKHorodecki}.  Entangled quantum states show
correlations between measurements on the subsystems which can not be
modelled within classical physics with local interactions.  A
classical model would have to be a joint probability distribution of
quantitities that are incompatible in the quantum theory, and the
existence of such a joint probability distribution, consistent with
locality, implies so called Bell inequalities~\cite{Bell64}, or even
equalities as in the three particle states known as GHZ states,
introduced by Mermin, Greenberger, Horne, and
Zeilinger~\cite{GHZ89,Mermin90}.  The correlations in entangled
quantum states violate Bell inequalities and GHZ equalities.

A pure classical state of a composite system has no correlations
between measurements on subsystems, since classical measurements are
deterministic.  A statistical ensemble of pure classical states, what
we may call a mixed classical state, can have correlations, but these
correlations can not violate Bell inequalities, by definition.

The only pure quantum states that are not entangled are the pure
product states, which resemble pure classical states in that they have
no correlations at all.  By definition, a mixed quantum state is a
statistical ensemble of pure quantum states, and it is said to be
separable if it can be mixed entirely from pure product states.  The
separable mixed states are not entangled, since they can not violate
Bell inequalities.  The entangled mixed states are precisely those
that are non-separable.  For this reason, the mathematical distinction
between separable and non-separable mixed states is important from the
physical point of view.

The separability problem, how to characterize the set $\cS$ of
separable mixed states and decide whether a given mixed state is
separable or not, is known to be a difficult mathematical
problem~\cite{Gurvits03}.  It motivates our work presented here and in
previous papers, although we have not studied so much the separable
states directly as the larger class of mixed states called PPT
states~\cite{LeinaasMyrheim06,LeinaasMyrheim07, Sollid10a,Sollid10b}.

The separable mixed states have the property that they remain positive
after partial transposition, they are PPT states, for short.  The set
$\cP$ of PPT states is in general larger than the set $\cS$ of
separable states, but the difference between the two sets is
surprisingly small in low dimensions, and in the very lowest
dimensions,
$2\times 2$, $2\times 3$, and $3\times 2$, there is no
difference~\cite{Horodecki96}.

The condition of positive partial transpose is known as the Peres
separability criterion~\cite{Peres96}.  It is a powerful separability
test, especially in low dimensions where the difference between the
two sets $\cP$ and $\cS$ is small.  It can be used for example to
prove that any pure quantum state is either entangled or a pure
product state.

We study especially the lowest rank entangled PPT states, on the
assumption that they are the easiest ones to understand.  In the
present paper we discuss in particular how to construct PPT states of
rank 4 and 5 in $3\times 3$ dimensions, and rank 6 in $4\times 4$
dimensions.  A central theme is how the states are constrained by the
existence of product vectors in the kernel.  Another central theme is
perturbation theory, which we use to contruct rank 5 PPT states close
to rank 4 states, and to study surfaces of PPT states of fixed low
rank.  We compute numerically the dimensions of such surfaces, and we
show how to follow a surface by numerical integration of an equation
of motion.

\subsubsection*{The relation between PPT states and product vectors}

The close connection between PPT states and product vectors has been
used earlier, for example to prove the separability of sufficiently
low rank PPT states~\cite{Horodecki00}.

Bennett et al.~\cite{Bennett99,DiVincenzo03} introduced a method for
constructing low rank mixed states that are obviously entangled PPT
states, using what they called Unextendible Product Bases (UPBs).
From a UPB, defined as a maximal set of orthogonal product vectors
which is not a complete basis of the Hilbert space, one constructs an
orthogonal projection $Q$ and the complementary projection
$P=\iden-Q$.  Then $\gr=P/(\Tr P)$ is an entangled PPT state.

The UPB construction is most successful in the special case of rank 4
PPT states in $3\times 3$ dimensions.  In Ref.~\cite{Sollid10b} we
argued, partly based on evidence from numerical studies, that an
extended version of the UPB construction, including nonunitary but
nonsingular product transformations on the states, is general enough
to produce all rank 4 entangled PPT states in $3\times 3$ dimensions.

Unfortunately, attempts to apply the UPB method directly in higher
dimensions fail, even when the kernel contains product vectors,
because there can not exist a sufficient number of orthogonal product
vectors.  The orthogonality is essential in the construction by
Bennett et al.\ of the PPT state as a projection operator.  We would
like to generalize the construction in such a way that it works
without the orthogonality condition.

One possible generalization is to construct projection operators as
more general convex combinations, or even as non-convex linear
combinations, of pure product states.  This idea is explored in a
separate paper~\cite{SollidLeinaas10}.

In the present paper we discuss in general the constraints imposed on
a PPT state $\gr$ by the existence of product vectors in its kernel,
and we show that these constraints are so strong that they actually
determine the state uniquely.  A surprising discovery is that in cases
where the kernel contains a finite overcomplete set of product
vectors, the state $\gr$ can be reconstructed from only a subset of
the product vectors, and the number of product vectors needed may even
be smaller than the dimension of the kernel.

From this point of view, the important question is exactly what
conditions the product vectors must satisfy in order for the
constraint equations to have a solution for $\gr$.  We can answer this
question in the familiar special case of rank 4 PPT states in
$3\times 3$ dimensions, but not in other cases.  We consider this an
interesting problem for future research.

\subsubsection*{Outline of the paper}

The contents of the present paper are organized as follows.  First we
review some linear algebra, in particular degenerate perturbation
theory, in Sections~\ref{secbasiclinalg}, \ref{secrestrpert}, and
\ref{secprodvecker}.  The main purpose is to introduce notation and
collect formulas for later reference.

In Section~\ref{secrank44ppt} we discuss the rank 4 PPT states in
$3\times 3$ dimensions.  We review the UPB construction, based on
orthogonal product vectors in the kernel, before we describe an
approach which is different in that the product vectors need not be
orthogonal.  The new approach also throws some new light on a set
of reality conditions that limit the selection of product vectors
to be used for constructing rank 4 PPT states.

In Section~\ref{secrank55ppt} we discuss the rank 5 PPT states in
$3\times 3$ dimensions.  We find an 8 dimensional surface of rank 5
PPT states in every generic 5 dimensional subspace, but we have not
found any general method to construct such states.  However, we show
how to construct rank 5 PPT states by perturbing rank 4 PPT states.
Again, the product vectors in the kernel of the rank 4 state play an
important role in our construction of the rank 5 states.

In Section~\ref{secrank66ppt} we discuss rank 6 PPT states in $4\times
4$ dimensions.  The kernel of such a state has dimension 10, and
contains 20 product vectors.  The remarkable result we find is that
the state can be constructed from only 7 product vectors in the
kernel.  An arbitrary set of 7 product vectors does not produce a rank
6 PPT state, but we do not know how to select sets of product vectors
that can be used in such a construction.

In Section~\ref{secdimcount} we discuss briefly how to determine
numerically the dimensions of surfaces of PPT states of fixed rank.
We find that the dimensions are given by a simple counting of
independent constraints, except for the very lowest rank states, for
which the constraints are not independent.

Finally, we discuss in Section~\ref{secnumint} how to study a surface
of PPT states by numerical integration of equations of motion for
curves on the surface.  In this way one may study for example the
curvature of the surface, or how a curve on the surface approaches the
boundary of the surface.


\section{Some basic linear algebra}
\label{secbasiclinalg}

\subsection{Density matrices}

Let $H_N$ be the set of Hermitean $N\times N$ matrices.  It has a
natural structure as a real Hilbert space of dimension $N^2$ with the
scalar product
\be{eq004}
(X,Y)=\Tr(XY)\;.
\ee

A mixed state, or density matrix, is a positive Hermitean matrix of
unit trace.  We define
\be{eq00db}
\cD=\cD_N=\{\,\gr\in H_N \mid \gr\geq 0\,,\,\Tr\gr=1\,\}\;.
\ee
%
%
Because it is Hermitean, a density matrix $\gr$ has a spectral
representation in terms of a complete set of orthonormal eigenvectors
$\psi_i\in\mathbbm{C}^N$ with real eigenvalues $\gl_i$,
\be{eq00a}
\gr=\sum_{i=1}^N \gl_i\,\psi_i{\psi_i}^{\!\dag}
\qquad\mbox{with}\qquad
{\psi_i}^{\!\dag}\psi_j=\gd_{ij}\;,\qquad
\iden=\sum_{i=1}^N \psi_i{\psi_i}^{\!\dag}\;.
\ee
%
The rank of $\gr$ is the number of eigenvalues $\gl_i\neq 0$.  The
pseudoinverse of $\gr$ is defined as
\be{eq00ba}
\gr^+=\sum_{i,\gl_i\neq 0} {\gl_i}^{-1}\,\psi_i{\psi_i}^{\!\dag}\;,
\ee
it is equal to the inverse $\gr^{-1}$ if $\gr$ is invertible.  The
matrices
\be{eq00b}
P=\gr^+\!\gr=\gr\gr^+=\sum_{i,\gl_i\neq 0} \psi_i{\psi_i}^{\!\dag}\;,\qquad
Q=\iden-P=\sum_{i,\gl_i=0} \psi_i{\psi_i}^{\!\dag}
\ee
are Hermitean and project orthogonally onto two complementary
orthogonal subspaces of $\mathbbm{C}^N$, $P$ onto $\Img\gr$, the image
of $\gr$, and $Q$ onto $\Ker\gr$, the kernel of $\gr$.  The relations
$P\gr=\gr P=P\gr P=\gr$ and $Q\gr=\gr Q=Q\gr Q=0$ will be used in the
following.

We say that $\gr$ is positive, or positive semidefinite, and we write
$\gr\geq 0$, when $\gl_i\geq 0$ for $i=1,2,\ldots,N$.  An equivalent
condition is that $\psi^{\dag}\gr\psi\geq 0$ for all
$\psi\in\mathbbm{C}^N$.  It follows from the last inequality and the
spectral representation of $\gr$ that $\psi^{\dag}\gr\psi=0$ if and
only if $\gr\psi=0$.

The definition of positive Hermitean matrices by inequalities of the
form $\psi^{\dag}\gr\psi\geq 0$ implies that $\cD$ is a convex set.
That is, if $\gr$ is a convex combination of $\gr_1,\gr_2\in\cD$,
\be{eq00f}
\gr=\gl\gr_1+(1-\gl)\gr_2\qquad\mbox{with}\qquad 0<\gl<1\;,
\ee
then $\gr\in\cD$.  Furthermore, since
$\Ker\gr=\{\psi\mid\psi^{\dag}\gr\psi=0\}$ when $\gr\geq 0$, it follows
that
\be{eq00g}
\Ker\gr=\Ker\gr_1\cap\Ker\gr_2\;,
\ee
independent of $\gl$, when $\gr$ is a convex combination as above.
Since $\Ker\gr$ is independent of $\gl$, so is
$\Img\gr=(\Ker\gr)^{\perp}$.

A convex set is defined by its extremal points: those points that are
not convex combinations of other points.  The extremal points of $\cD$
are the pure states of the form $\gr=\psi\psi^{\dag}$ with
$\psi\in\mathbbm{C}^N$.  Thus, the spectral representation in
eq.~(\ref{eq00a}) is an expansion of $\gr$ as a convex combination of
$N$ or fewer extremal points of $\cD$.

\subsubsection*{Finite perturbations}

In the following, let $\gr$ be a density matrix and define the
projections $P$ and $Q=\iden-P$ as in eq.~(\ref{eq00b}).  Consider a
perturbation
\be{eq001}
\gr'=\gr+\ge A\;,
\ee
where $A\neq 0$ is Hermitean, and $\Tr A=0$ so that $\Tr\gr'=\Tr\gr$.
The real parameter $\ge$ may be finite or infinitesimal, we will first
consider the case when $\ge$ is finite.

We observe that if $\Img A\subset\Img\gr$, or equivalently if $PAP=A$,
then there will be a finite range of values of $\ge$, say
$\ge_1\leq\ge\leq\ge_2$ with $\ge_1<0<\ge_2$, such that $\gr'\in\cD$
and $\Img\gr'=\Img\gr$.  This is so because the eigenvectors of $\gr$
with zero eigenvalue will remain eigenvectors of $\gr'$ with zero
eigenvalue, and all the positive eigenvalues of $\gr$ will change
continuously with $\ge$ into eigenvalues of $\gr'$.

The other way around, if $\gr'\in\cD$ for $\ge_1\leq\ge\leq\ge_2$ with
$\ge_1<0<\ge_2$, then $\gr'$ is a convex combination of $\gr+\ge_1 A$
and $\gr+\ge_2 A$ for every $\ge$ in the open interval
$\ge_1<\ge<\ge_2$.  Hence $\Img\gr'$ is independent of $\ge$ in this
interval, implying that $\Img A\subset\Img\gr$ and $PAP=A$.

This shows that $\gr$ is extremal in $\cD$ if and only if there exists
no $A\neq 0$ with $\Tr A=0$ and $PAP=A$.  Another formulation of the
condition is that there exists no $\gr'\in\cD$ with $\gr'\neq\gr$ and
$\Img\gr'=\Img\gr$.  A third equivalent formulation of the extremality
condition is that the equation
\be{eq00ha}
PAP=A
\ee
for the Hermitean matrix $A$ has $A=\gr$ as its only solution (up to
proportionality).  In fact, if $PBP=B$ and $\Tr B\neq 0$, then we have
$PAP=A$ and $\Tr A=0$ when we take
\be{eq00i}
A=B-(\Tr B)\,\gr\;.
\ee

\subsubsection*{Infinitesimal perturbations}

Assume now that $\Img A\not\subset\Img\gr$.  The question is how an
infinitesimal perturbation affects the zero eigenvalues of $\gr$.
When $\gr$ is of low rank we need degenerate perturbation theory,
which is well known from any textbook on quantum mechanics.

To first order in $\ge$, the zero eigenvalues of $\gr$ are perturbed
into eigenvalues of $\gr'$ that are $\ge$ times the eigenvalues of
$QAQ$ on the subspace $\Ker\gr$.  Similarly, to first order in $\ge$,
the positive eigenvalues of $\gr$ are perturbed into positive
eigenvalues of $\gr'$, in a way which is determined by how $\gr$ and
$PAP$ act on $\Img\gr$.

It is clear from this that, to first order in $\ge$, the condition
\be{eq00ia}
QAQ=0
\ee
is necessary and sufficient to ensure that the rank of $\gr'$ equals
the rank of $\gr$, and that $\gr'\geq 0$ both for $\ge>0$ and for
$\ge<0$.

More generally, to first order in $\ge$, the rank of $\gr'$ equals the
rank of $\gr$ plus the rank of $QAQ$.  For example, if we want to
perturb $\gr$ in such a way that the rank increases by one, then we
have to choose $A$ such that
\be{eq00j}
QAQ=\ga\,\phi\phi^{\dag}\;,
\ee
where $\phi\in\Ker\gr$ is a normalized eigenvector of $QAQ$ with
$\ga\neq 0$ as eigenvalue.  Since $QAQ$ is Hermitean, $\ga$ must be
real.  If $\ga>0$, then $\gr'\geq 0$ for $\ge>0$ but not for $\ge<0$.

\subsubsection*{Projection operators on $H_N$}

Using the projections $P$ and $Q$ defined above we define projection
operators on $H_N$, the real Hilbert space of Hermitean $N\times N$
matrices, as follows,
\be{eq002}
{\rm\bf P}X
&\!\!\!=&\!\!\!PXP\;,\nonumber\\
{\rm\bf Q}X&\!\!\!=&\!\!\!QXQ=X-PX-XP+PXP\;,\\
{\rm\bf R}X&\!\!\!=&\!\!\!({\rm\bf I}-{\rm\bf P}-{\rm\bf Q})X
=PX+XP-2PXP\;.\nonumber
\ee
Here ${\rm\bf I}$ is the identity on $H_N$.  It is straightforward to
verify that these are complementary projections, with
%
${\rm\bf P}^2={\rm\bf P}$,
${\rm\bf Q}^2={\rm\bf Q}$,
${\rm\bf P}{\rm\bf Q}={\rm\bf Q}{\rm\bf P}={\bf 0}$,
and so on.
%
They are symmetric with respect to the natural scalar product on
$H_N$, for example,
\be{eq005}
(X,{\rm\bf P}Y)=\Tr(XPYP)=\Tr(PXPY)=({\rm\bf P}X,Y)\;.
\ee
Hence they project orthogonally,
and relative to an orthonormal basis of $H_N$ they are represented by
symmetric matrices.

Relative to an orthonormal basis of $\mathbbm{C}^N$ with the first
basis vectors in $\Img\gr$ and the last basis vectors in $\Ker\gr$, a
Hermitean matrix $X$ takes the block form
\be{eq006}
X=\left(\begin{array}{cc}
U&V\\ V^{\dag}&W
\end{array}\right)
\;,
\ee
with $U^{\dag}=U$ and $W^{\dag}=W$.  In this basis we have
\be{eq007}
P=\left(\begin{array}{cc}
I&0\\ 0&0
\end{array}\right)
\;,\qquad
Q=\left(\begin{array}{cc}
0&0\\ 0&I
\end{array}\right)
\;,
\ee
and hence,
\be{eq008}
{\rm\bf P}X=\left(\begin{array}{cc}
U&0\\ 0&0
\end{array}\right)
\;,\qquad
{\rm\bf Q}X=\left(\begin{array}{cc}
0&0\\ 0&W
\end{array}\right)
\;,\qquad
{\rm\bf R}X=\left(\begin{array}{cc}
0&V\\ V^{\dag}&0
\end{array}\right)
\;.
\ee

\subsection{Composite systems}

\subsubsection*{Product vectors}

If $N=N_AN_B$ then the tensor product spaces
$\mathbbm{C}^N=\mathbbm{C}^{N_A}\otimes\mathbbm{C}^{N_B}$ (a complex
tensor product) and $H_N=H_{N_A}\otimes H_{N_B}$ (a real tensor
product) may describe a composite quantum system with two subsystems
$A$ and $B$ of Hilbert space dimensions $N_A$ and $N_B$.

A vector $\psi\in\mathbbm{C}^N$ then has components
$\psi_I=\psi_{ij}$, where
\be{eq00k}
I=1,2,\ldots,N\quad\leftrightarrow\quad
ij=11,12,\ldots,1N_B,21,22,\ldots,N_AN_B\;.
\ee
A product vector $\psi=\phi\otimes\chi$ has components
$\psi_{ij}=\phi_i\chi_j$.  We see that $\psi$ is a product vector if
and only if its components satisfy the quadratic equations
\be{eq00ka}
\psi_{ij}\psi_{kl}-\psi_{il}\psi_{kj}=0\;.
\ee
These equations are not all independent, the number of independent
complex equations is
\be{eq00kaa}
K=(N_A-1)(N_B-1)=N-N_A-N_B+1\;.
\ee
For example, if $\psi_{11}\neq 0$ we get a complete set of independent
equations by taking $i=j=1$ and $k=2,3,\ldots,N_A$,
$l=2,3,\ldots,N_B$.

Since the equations are homogeneous, any solution $\psi\neq 0$ gives
rise to a one parameter family of solutions $c\psi$ with
$c\in\mathbbm{C}$.  A vector $\psi$ in a subspace of dimension $n$ has
$n$ independent complex components.  Since the most general nonzero
solution must contain at least one free complex parameter, we conclude
that a generic subspace of dimension $n$ will contain nonzero product
vectors if and only if
\be{eq00kb}
n\geq K+1=N-N_A-N_B+2\;.
\ee
The limiting dimension
\be{eq00kc}
n=N-N_A-N_B+2
\ee
is particularly interesting.  In this special case a nonzero solution
will contain exactly one free parameter, which has to be a complex
normalization constant.

Thus, up to proportionality there will exist a finite set of product
vectors in a generic subspace of dimension $n=N-N_A-N_B+2$.
The number of product vectors is~\cite{Sollid10a}
\be{eq00kd}
p=\pmatrix{N_A+N_B-2 \cr N_A-1}
=\frac{(N_A+N_B-2)!}{(N_A-1)!\,(N_B-1)!}\;.
\ee
A generic subspace of lower dimension will contain no nonzero product
vector, whereas any subspace of higher dimension will contain a
continuous infinity of different product vectors (different in the
sense that they are not proportional).

\subsubsection*{Partial transposition}

The following relation between matrix elements,
\be{eq00l}
(X^P)_{ij;kl}=X_{il;kj}\;,
\ee
defines $X^P$, the partial transpose of the Hermitean matrix $X$ with
respect to the second subsystem.

A density matrix $\gr$ is called separable if it is a convex
combination of tensor product matrices,
\be{eq00da}
\gr=\sum_k p_k\,\gs_k\otimes\gt_k\;,
\ee
with $\gs_k\in\cD_{N_A}$, $\gt_k\in\cD_{N_B}$, $p_k>0$, $\sum_kp_k=1$.
We denote by $\cS$ the set of separable matrices.

The partial transpose of the above separable matrix is
\be{eq00dc}
\gr^P=\sum_k p_k\,\gs_k\otimes(\gt_k)^T\geq 0\;.
\ee
The positivity of $\gr^P$ is known as the Peres criterion, it is an
easily testable necessary condition for separability.  For this reason
it is of interest to study the set of PPT (Positive Partial Transpose)
matrices, defined as
\be{eq00m}
\cP=\{\,\gr\in\cD \mid \gr^P\geq 0\,\}=\cD\cap\cD^P\;.
\ee
We may call it the Peres set.  A well known result is that $\cP=\cS$
for $N=N_AN_B\leq 6$, whereas $\cP$ is strictly larger than $\cS$ in
higher dimensions~\cite{Horodecki96}.

We will classify low rank PPT states by the ranks $(m,n)$ of $\gr$ and
$\gr^P$, respectively.  Note that ranks $(m,n)$ and $(n,m)$ are
equivalent for the purpose of classification, because of the symmetric
roles of $\gr$ and $\gr^P$.

\subsubsection*{Product transformations}

A product transformation of the form
\be{eq00ma}
\gr\mapsto\gr'=aV\gr V^{\dag}
\qquad\mbox{with}\qquad
V=V_A\otimes V_B\;,
\ee
where $a$ is a normalization factor and
$V_A\in\mbox{SL}(N_A,\mathbbm{C})$,
$V_B\in\mbox{SL}(N_B,\mathbbm{C})$, preserves positivity, rank,
separability, and other interesting properties that the density matrix
$\rho$ may have.  For example, it preserves positivity of the partial
transpose, because
\be{eq00mb}
(\gr')^P=a\widetilde{V}\gr^P\widetilde{V}^{\dag}
\qquad\mbox{with}\qquad
\widetilde{V}=V_A\otimes V_B^{\;\ast}\;.
\ee
The image and kernel of $\gr$ and $\gr^P$ transform in the following
ways,
\be{eq00mc}
\Img\gr'=V\Img\gr\;,\qquad
\Ker\gr'=(V^{\dag})^{-1}\Ker\gr\;, \ee
and
\be{eq00md}
\Img\,(\gr')^P=\widetilde{V}\Img\gr^P\;,\qquad
\Ker\,(\gr')^P=(\widetilde{V}^{\dag})^{-1}\Ker\gr^P\;.
\ee
All these transformations are of product form and hence preserve the
number of product vectors in a subspace.

We say that two density matrices $\gr$ and $\gr'$ related in this way
are $\mbox{SL}\otimes\mbox{SL}$ equivalent, or simply $\mbox{SL}$
equivalent.  The concept of $\mbox{SL}$ equivalence is important to us
here because it simplifies very much our efforts to classify the low
rank PPT states.

\section{Restricted perturbations}
\label{secrestrpert}

We have seen that eq.~(\ref{eq00ha}) ensures that the perturbation
$\gr'=\gr+\ge A$ preserves the image of $\gr$, so that
$\Img\gr'=\Img\gr$ for infinitesimal values of the perturbation
parameter $\ge$, and $\Img\gr'\subset\Img\gr$ even for finite values
of $\ge$.  The weaker condition in eq.~(\ref{eq00ia}) ensures only
that the rank of $\gr'$ equals the rank of $\gr$ for infinitesimal
values of $\ge$.

We want to discuss how to use perturbations with similar restrictions
in order to study, for example, the extremal points of the convex set
$\cP$.  In particular, we are interested in perturbations that either
preserve the ranks $(m,n)$ of $\gr$, or else change these ranks in
controlled ways.

In a similar way as we did for $\gr$, we define $\wt{P}$ and
$\wt{Q}=\iden-\wt{P}$ as the orthogonal projections onto $\Img\gr^P$
and $\Ker\gr^P$.  Then we define
\be{eq009}
\wt{\rm\bf P}X
&\!\!\!=&\!\!\!(\wt{P}X^P\wt{P})^P\;,\nonumber\\
\wt{\rm\bf Q}X
&\!\!\!=&\!\!\!(\wt{Q}X^P\wt{Q})^P
=X-(\wt{P}X^P)^P-(X^P\wt{P})^P
+(\wt{P}X^P\wt{P})^P\;,\\
\wt{\rm\bf R}X
&\!\!\!=&\!\!\!({\rm\bf I}-\wt{\rm\bf P}-\wt{\rm\bf Q})X
=(\wt{P}X^P)^P+(X^P\wt{P})^P
-2(\wt{P}X^P\wt{P})^P\;.\nonumber
\ee
These are again projections on the real Hilbert space $H_N$, like
${\rm\bf P}$, ${\rm\bf Q}$ and ${\rm\bf R}$, again symmetric with
respect to the natural scalar product on $H_N$.

We may now use the projection operators on $H_N$ to impose various
restrictions on the perturbation matrix $A$.

\subsubsection*{Testing for extremality in $\cP$}

The extremality condition for $\cP$ is derived in a similar way as the
extremality condition for $\cD$ based on eq.~(\ref{eq00ha}).  Clearly
$\gr$ is extremal in $\cP$ if and only if there exists no
$\gr'\in\cP$, $\gr'\neq\gr$, with both $\Img\gr'=\Img\gr$ and
$\Img(\gr')^P=\Img\gr^P$.  Another way to formulate this condition is
that $A=\gr$ is the only solution of the two equations ${\rm\bf P}A=A$
and $\wt{\rm\bf P}A=A$.

Since ${\rm\bf P}$ and $\wt{\rm\bf P}$ are projections, the equations
${\rm\bf P}A=A$ and $\wt{\rm\bf P}A=A$ together are equivalent to the
single eigenvalue equation
\be{eq010}
({\rm\bf P}+\wt{\rm\bf P})A=2A\;.
\ee
They are also equivalent to either one of the eigenvalue equations
\be{eq011}
   {\rm\bf P}\wt{\rm\bf P}   {\rm\bf P}A=A\;,\qquad
\wt{\rm\bf P}   {\rm\bf P}\wt{\rm\bf P}A=A\;.
\ee
Note that the operators ${\rm\bf P}+\wt{\rm\bf P}$,
${\rm\bf P}\wt{\rm\bf P}{\rm\bf P}$, and
$\wt{\rm\bf P}{\rm\bf P}\wt{\rm\bf P}$ are all real symmetric and
therefore have complete sets of real eigenvalues and eigenvectors.  In
fact, the eigenvalues are all non-negative, because the operators are
positive semidefinite.

When we diagonalize ${\rm\bf P}+\wt{\rm\bf P}$ we will always find
$A=\gr$ as an eigenvector with eigenvalue 2.  If it is the only
solution of eq.~(\ref{eq010}), this proves that $\gr$ is extremal in
$\cP$.  If $A$ is a solution not proportional to $\gr$, then we may
impose the condition $\Tr A=0$ (replace $A$ by $A-(\Tr A)\gr$ if
necessary), and we know that there exists a finite range of both
positive and negative values of $\ge$ such that
$\gr'=\gr+\ge A\in\cP$, hence $\gr$ is not extremal.

It should be noted that in our numerical calculations we may find
eigenvalues of ${\rm\bf P}+\wt{\rm\bf P}$ that differ from 2 by less
than one per cent.  However, since eigenvalues are calculated with a
precision close to the internal precision of the computer, which is of
order $10^{-16}$, there is never any ambiguity as to whether an
eigenvalue is equal to 2 or strictly smaller than 2.

\subsubsection*{Perturbations preserving the PPT property and ranks}

The rank and positivity of $\gr$ is preserved by the perturbation, to
first order in $\ge$, both for $\ge>0$ and $\ge<0$, if and only if
${\rm\bf Q}A=0$.  Similarly, the rank and positivity of $\gr^P$ is
preserved if and only if
$\wt{\rm\bf Q}A=0$.  These two equations together are equivalent to
the single eigenvalue equation
\be{eq012}
({\rm\bf Q}+\wt{\rm\bf Q})A=0\;.
\ee
Again ${\rm\bf Q}+\wt{\rm\bf Q}$ is real symmetric and has a complete
set of real eigenvalues and eigenvectors.

In conclusion, the perturbations that preserve the PPT property, as
well as the ranks $(m,n)$ of $\gr$ and $\gr^P$, to first order in
$\ge$, are the solutions of eq.~(\ref{eq012}).

We may want to perturb in different ways, for example such that
$\Img\gr'=\Img\gr$, but not necessarily $\Img(\gr')^P=\Img\gr^P$, we
only require $(\gr')^P$ and $\gr^P$ to have the same rank.  Then the
conditions on $A$ are that ${\rm\bf P}A=A$ and $\wt{\rm\bf Q}A=0$, or
equivalently,
\be{eq013}
({\rm\bf I}-{\rm\bf P}+\wt{\rm\bf Q})A=0\;.
\ee
%

\section{Product vectors in the kernel}
\label{secprodvecker}

Assume that $\gr\in\cP$.  Recall that the equations $w^{\dag}\gr w=0$
and $\gr w=0$ for $w\in\mathbbm{C}^N$ are equivalent, and so are the
equations $w^{\dag}\gr^Pw=0$ and $\gr^Pw=0$, because $\gr\geq 0$ and
$\gr^P\geq 0$.  Taken together with the identity
\be{eqmatelrhorhoP}
(x\otimes y)^{\dag}\gr\,(u\otimes v)=
(x\otimes v^{\ast})^{\dag}\gr^P(u\otimes y^{\ast})
\ee
this puts strong restrictions on $\gr$ when we know a number of
product vectors in $\Ker\gr$.

Assume from now on that $w$ is a product vector, $w=u\otimes v$.
Defining $\widetilde{w}=u\otimes v^{\ast}$ we have the general
relation
\be{eqwrhowwtrhoPwt}
w^{\dag}\gr w=\widetilde{w}^{\,\dag}\gr^P\widetilde{w}\;.
\ee
Assume furthermore that $w\in\Ker\gr$, this is equivalent to the
condition that $\widetilde{w}\in\Ker\gr^P$.  For any
$z\in\mathbbm{C}^N$ we have the condition on $\gr$ that $z^{\dag}\gr
w=0$.  In particular, when $z$ is an arbitrary product vector,
$z=x\otimes y$, we have the two conditions on $\gr$ that
\be{eq079}
(x\otimes y)^{\dag}\gr\,(u\otimes v)=0
\ee
and
\be{eq080}
(x\otimes v)^{\dag}\gr\,(u\otimes y)=
(x\otimes y^{\ast})^{\dag}\gr^P(u\otimes v^{\ast})=0\;.
\ee

Assume that $w_i=u_i\otimes v_i\in\Ker\gr$ for $i=1,2,\ldots,n$.  Then
for arbitrary values of the indices $i,j,k$ we have the following
constraints on $\gr$,
\be{eq081}
(u_i\otimes v_j)^{\dag}\gr\,(u_k\otimes v_k)=
(u_i\otimes v_k)^{\dag}\gr\,(u_k\otimes v_j)=0\;.
\ee
Let us introduce matrices
\be{eq082}
A_{klij} = (u_k\otimes v_l)(u_i\otimes v_j)^{\dag}\;,
\ee
and Hermitean matrices
\be{eq083}
B_{klij}=A_{klij}+(A_{klij})^{\dag}\;,\qquad
C_{klij}={\rm i}\,(A_{klij}-(A_{klij})^{\dag})\;,
\ee
then the constraints on $\gr$ are of the form
\be{eq084}
\Tr(\gr B_{kkij})=
\Tr(\gr C_{kkij})=
\Tr(\gr B_{kjik})=
\Tr(\gr C_{kjik})=0\;.
\ee
Each equation $\Tr(\gr B)=0$ with $B\neq 0$ or $\Tr(\gr C)=0$ with
$C\neq 0$ is one real valued constraint.  Of course, the constraints
in eq.~(\ref{eq084}) are not all independent, we have for example that
$C_{kkkk}=0$.

\section{Rank $(4,4)$ PPT states in $3\times 3$ dimensions}
\label{secrank44ppt}

\subsection{The UPB construction of entangled PPT states}
\label{subsecUPB}

We will review briefly the construction of a rank $(4,4)$ entangled
and extremal PPT state $\gr$ in $3\times 3$ dimensions from an
unextendible orthonormal product basis (a UPB) of $\Ker\rho$
\cite{Bennett99}.  The UPB consists of five orthonormal product
vectors $w_i=N_iu_i\otimes v_i$ with the property that there exists no
product vector orthogonal to all of them.  We include real
normalization factors $N_i$ here because we want to normalize such
that $w_i^{\,\dag}w_j=\gd_{ij}$ without necessarily normalizing the
vectors $u_i$ and $v_i$.

The orthogonality of the product vectors $w_i$ follows from the
orthogonality relations
$u_1\perp u_2\perp u_3\perp u_4\perp u_5\perp u_1$ and
$v_1\perp v_3\perp v_5\perp v_2\perp v_4\perp v_1$.  There is the
further condition that any three vectors $u_i$ and any three $v_i$ are
linearly independent.  The five dimensional subspace spanned by
these product vectors is the kernel of the density matrix
\be{eq085}
\gr = \frac{1}{4}\left(\iden-\sum_{i=1}^5w_iw_i^{\,\dag}\right),
\ee
which is proportional to a projection operator.  The partial transpose
of $\gr$ is
\be{eq086}
\gr^P = \frac{1}{4}\left(\iden-
\sum_{i=1}^5\widetilde{w}_i\widetilde{w}_i^{\,\dag}\right),
\ee
with $\widetilde{w}_i=N_iu_i\otimes v_i^{\,\ast}$.  Thus we have both
$\gr\geq 0$ and $\gr^P\geq 0$ by construction.  Note that if all the
vectors $v_i$ are real, $v_i^{\,\ast}=v_i$, then $\gr$ is symmetric
under partial transposition, $\gr^P=\gr$.

By a unitary product transformation as in eq.~(\ref{eq00mc}) we may
transform the above orthogonal UPB into the standard unnormalized form
\cite{Sollid10b}
\be{equvorth}
u=\pmatrix{
1 & 0 & a & b & 0\cr
0 & 1 & 0 & 1 & a\cr
0 & 0 & b &-a & 1},
\qquad
v=\pmatrix{
1 & d & 0 & 0 & c\cr
0 & 1 & 1 & c & 0\cr
0 &-c & 0 & 1 & d},
\ee
with $a,b,c,d$ as positive real parameters.  The following quantities
determine these parameters,
\be{inv12}
s_1&\!\!\!=&\!\!\!
-\frac{\det(u_1 u_2 u_4)\,\det(u_1 u_3 u_5)}
      {\det(u_1 u_2 u_5)\,\det(u_1 u_3 u_4)}=a^2\;,
\nonumber\\
s_2&\!\!\!=&\!\!\!
-\frac{\det(u_1 u_2 u_3)\,\det(u_2 u_4 u_5)}
      {\det(u_1 u_2 u_4)\,\det(u_2 u_3 u_5)}=\frac{b^2}{a^2}\;,
\ee
and
\be{inv34}
s_3&\!\!\!=&\!\!\!
\frac{\det(v_1 v_2 v_3)\,\det(v_1 v_4 v_5)}
     {\det(v_1 v_2 v_5)\,\det(v_1 v_3 v_4)}=c^2\;,
\nonumber\\
s_4&\!\!\!=&\!\!\!
\frac{\det(v_1 v_3 v_5)\,\det(v_2 v_3 v_4)}
     {\det(v_1 v_2 v_3)\,\det(v_3 v_4 v_5)}=\frac{d^2}{c^2}\;.
\ee
These ratios of determinants are invariant under general
$\mbox{SL}\otimes\mbox{SL}$ transformations, as well as independent of
the normalization of the vectors.

In Ref.~\cite{Sollid10b} we presented numerical evidence that every
entangled rank $(4,4)$ PPT state is $\mbox{SL}\otimes\mbox{SL}$
equivalent to some state of the form of eq.~(\ref{eq085}) with real
product vectors as given in eq.~(\ref{equvorth}).

This means that the surface of all rank $(4,4)$ entangled PPT states
has dimension 36.  We count 32 degrees of freedom due to the
$\mbox{SL}(3,\mathbbm{C})\otimes\mbox{SL}(3,\mathbbm{C})$
transformations, plus the 4 real $\mbox{SL}\otimes\mbox{SL}$ invariant
parameters $a,b,c,d$ in eq.~(\ref{equvorth}).

\subsection{A different point of view}

We will present here the construction of an entangled PPT state $\gr$
of rank $(4,4)$ as seen from a different point of view.  When $\gr$
has rank 4 it means that $\Ker\gr$ has dimension 5.  A generic 5
dimensional subspace in
$\mathbbm{C}^9=\mathbbm{C}^3\otimes\mathbbm{C}^3$ has a basis of
product vectors.  In fact, it contains exactly 6 product vectors, any
5 of which are linearly independent.  By eq.~(\ref{eq00kc}), 5 is the
limiting dimension for which the number of product vectors is nonzero
and finite, and the number 6 is consistent with eq.~(\ref{eq00kd}).

Any set of 5 product vectors $w_i=u_i\otimes v_i$, orthogonal or not,
may be transformed by an $\mbox{SL}\otimes\mbox{SL}$ transformation to
the standard unnormalized form
\be{equvnonorth}
u=\pmatrix{
1 & 0 & 0 & 1 & 1\cr
0 & 1 & 0 & 1 & p\cr
0 & 0 & 1 & 1 & q},
\qquad
v=\pmatrix{
1 & 0 & 0 & 1 & 1\cr
0 & 1 & 0 & 1 & r\cr
0 & 0 & 1 & 1 & s},
\ee
with $p,q,r,s$ as real or complex parameters.  We impose here again
the condition that any three $u_i$ and any three $v_i$ should be
linearly independent.  There is always a 6th product vector which is
a linear combination of the above 5,
\renewcommand{\arraystretch}{1.6}
\be{equvvecI}
u_6
=\left(\!\begin{array}{c}
\frac{\mbox{\normalsize $s-r$}}{\mbox{\normalsize $ps-qr$}}\\
\frac{\mbox{\normalsize $1-s$}}{\mbox{\normalsize $q-s$}}\\
\frac{\mbox{\normalsize $r-1$}}{\mbox{\normalsize $r-p$}}
\end{array}\!\right),
\qquad
v_6
=\left(\!\begin{array}{c}
\frac{\mbox{\normalsize $p-q$}}{\mbox{\normalsize $ps-qr$}}\\
\frac{\mbox{\normalsize $q-1$}}{\mbox{\normalsize $q-s$}}\\
\frac{\mbox{\normalsize $1-p$}}{\mbox{\normalsize $r-p$}}
\end{array}\!\right).
\ee
\renewcommand{\arraystretch}{1.0}

The values of the above invariants as functions of the new parameters
are
\be{inv1234}
s_1=-\frac{p}{q}\;,
\qquad
s_2=q-1\;,
\qquad
s_3=\frac{r-s}{s}\;,
\qquad
s_4=\frac{r}{1-r}\;.
\ee
The parameters $p,q,r,s$ are actually new invariants, they can not be
changed by $\mbox{SL}\otimes\mbox{SL}$ transformations.

If the values of $p,q,r,s$ are such that the invariants
$s_1,s_2,s_3,s_4$ are all real and strictly positive, then we may use
eq.~(\ref{inv12}) and eq.~(\ref{inv34}) to find corresponding values
of $a,b,c,d$, and we may transform from the non-orthogonal standard
form in eq.~(\ref{equvnonorth}) to the orthogonal standard form in
eq.~(\ref{equvorth}), which in turn defines the rank $(4,4)$ state
in eq.~(\ref{eq085}).

\begin{figure}[htp]

\begin{picture}(300,200)(24,20)

\setlength{\unitlength}{0.7cm}
\linethickness{1pt}

\thicklines

\put(2.5,5){\vector(1,0){8}}
\put(5,2.5){\vector(0,1){7.5}}

\thinlines

\put(2.7,2.7){\line(1,1){7.1}}
\put(7.5,2.5){\line(0,1){7.5}}
\put(2.5,7.5){\line(1,0){7.5}}

\put(4.35,3.35){\makebox(0,0){\bf{\sf{\Large{3}}}}}
\put(3.35,4.35){\makebox(0,0){\bf{\sf{\Large{5}}}}}
\put(3.85,6.25){\makebox(0,0){\bf{\sf{\Large{6}}}}}
\put(3.85,8.7){\makebox(0,0){\bf{\sf{\Large{1}}}}}
\put(6.25,3.85){\makebox(0,0){\bf{\sf{\Large{4}}}}}
\put(6.25,8.7){\makebox(0,0){\bf{\sf{\Large{12}}}}}
\put(8.75,3.85){\makebox(0,0){\bf{\sf{\Large{2}}}}}
\put(8.75,6.25){\makebox(0,0){\bf{\sf{\Large{8}}}}}
\put(6.8,5.8){\makebox(0,0){\bf{\sf{\Large{10}}}}}
\put(5.8,6.8){\makebox(0,0){\bf{\sf{\Large{9}}}}}
\put(9.25,8.25){\makebox(0,0){\bf{\sf{\Large{7}}}}}
\put(8.25,9.25){\makebox(0,0){\bf{\sf{\Large{11}}}}}

\put(10.9,7.5){\makebox(0,0){\bf{\sf{\large{$q=1$}}}}}
\put(7.5,2.2){\makebox(0,0){\bf{\sf{\large{$p=1$}}}}}
\put(10.9,5){\makebox(0,0){\bf{\Large{\sf{p}}}}}
\put(5,10.5){\makebox(0,0){\bf{\Large{\sf{q}}}}}

\linethickness{1pt}

\thicklines

\put(13.5,5){\vector(1,0){8}}
\put(16,2.5){\vector(0,1){7.5}}

\thinlines

\put(13.7,2.7){\line(1,1){7.1}}
\put(18.5,2.5){\line(0,1){7.5}}
\put(13.5,7.5){\line(1,0){7.5}}

\put(15.35,3.35){\makebox(0,0){\bf{\sf{\Large{8}}}}}
\put(14.35,4.35){\makebox(0,0){\bf{\sf{\Large{12}}}}}
\put(14.85,6.25){\makebox(0,0){\bf{\sf{\Large{11}}}}}
\put(14.85,8.7){\makebox(0,0){\bf{\sf{\Large{10}}}}}
\put(17.25,3.85){\makebox(0,0){\bf{\sf{\Large{7}}}}}
\put(17.25,8.7){\makebox(0,0){\bf{\sf{\Large{5}}}}}
\put(19.75,3.85){\makebox(0,0){\bf{\sf{\Large{9}}}}}
\put(19.75,6.25){\makebox(0,0){\bf{\sf{\Large{3}}}}}
\put(17.8,5.8){\makebox(0,0){\bf{\sf{\Large{1}}}}}
\put(16.8,6.8){\makebox(0,0){\bf{\sf{\Large{2}}}}}
\put(20.25,8.25){\makebox(0,0){\bf{\sf{\Large{4}}}}}
\put(19.25,9.25){\makebox(0,0){\bf{\sf{\Large{6}}}}}

\put(21.9,7.5){\makebox(0,0){\bf{\sf{\large{$s=1$}}}}}
\put(18.5,2.2){\makebox(0,0){\bf{\sf{\large{$r=1$}}}}}
\put(21.9,5){\makebox(0,0){\bf{\Large{\sf{r}}}}}
\put(16,10.5){\makebox(0,0){\bf{\Large{\sf{s}}}}}

\end{picture}

\caption{Regions for the parameters $p,q,r,s$ defined in
  eq.~(\ref{equvnonorth}) such that the product vectors
  $u_i\otimes v_i$ for $i=1,2,\ldots,5$ are in the kernel of a rank
  $(4,4)$ extremal PPT state.  The $(p,q)$ plane is divided into 12
  regions, with 12 corresponding regions in the $(r,s)$ plane.  The
  numbers 1 to 12 refer to the permutations of product vectors given
  in eq.~(\ref{eq087c}).  For example, if $(p,q)$ is in region 7,
  $(r,s)$ must also be in region 7.}

\label{fig1}

\end{figure}
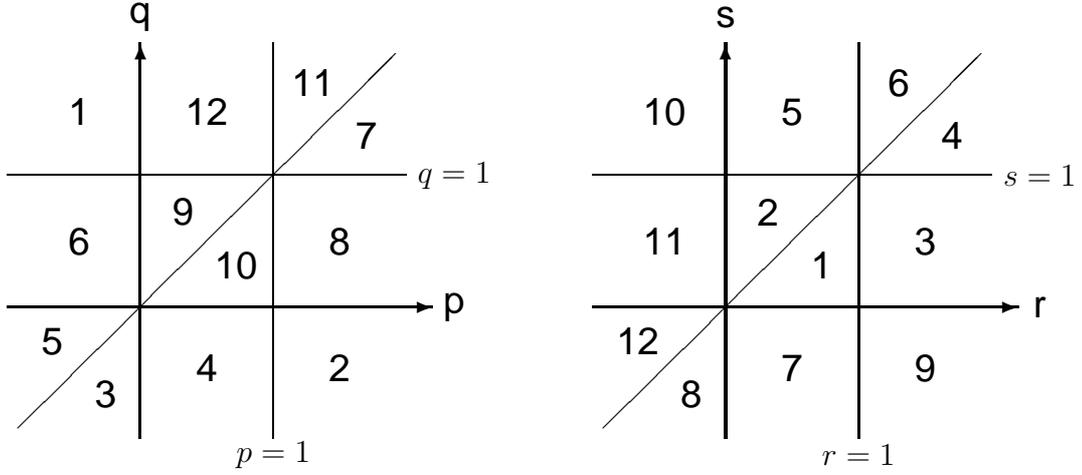

We see from eq.~(\ref{inv1234}) that the invariants are all strictly
positive if and only if $p,q,r,s$ are all real, and $p<0$, $q>1$,
$0<r<1$, $0<s<r$.  These inequalities define the regions marked 1 in
the $(p,q)$ and $(r,s)$ planes as plotted in Fig.~\ref{fig1}.

As discussed in ref.~\cite{Sollid10b} there are 10 permutations of the
product vectors $w_i$ for $i=1,2,\ldots,5$ which preserve the
positivity of the invariants.  These permutations form a group $G$
which is the symmetry group of a regular pentagon, exemplified by the
rotation, or cyclic permutation, $w_i\mapsto\widetilde{w}_i$ with
\be{eq087a}
\widetilde{w}_1=w_5\;,\quad
\widetilde{w}_2=w_1\;,\quad
\widetilde{w}_3=w_2\;,\quad
\widetilde{w}_4=w_3\;,\quad
\widetilde{w}_5=w_4\;,
\ee
and the reflection
\be{eq087b}
\widetilde{w}_1=w_4\;,\quad
\widetilde{w}_2=w_3\;,\quad
\widetilde{w}_3=w_2\;,\quad
\widetilde{w}_4=w_1\;,\quad
\widetilde{w}_5=w_5\;.
\ee
For short, we write the rotation as $51234$ and the reflection as
$43215$.

There are altogether $5!=120$ permutations of the 5 product vectors
$w_i$, and they fall into 12 classes (left cosets of the group $G$ as
a subgroup of the permutation group $S_5$) which are not transformed
into each other by $G$.  We number the classes from 1 to 12, and pick
one representative from each class as follows,
\be{eq087c}
\begin{array}{llllll}
\phantom{1}1:\; 12345 &
\phantom{1}2:\; 13245 & 
\phantom{1}3:\; 21345 & 
\phantom{1}4:\; 23145 & 
\phantom{1}5:\; 31245 & 
\phantom{1}6:\; 32145\\ 
\phantom{1}7:\; 12435 & 
\phantom{1}8:\; 14235 & 
\phantom{1}9:\; 21435 & 
          10:\; 24135 & 
          11:\; 13425 & 
          12:\; 14325
\end{array}
\ee
Each of these 12 classes defines a positivity region in each of the
two parameter planes, where all 4 invariants
$\wt{s}_1,\wt{s}_2,\wt{s}_3,\wt{s}_4$ computed from the permuted
product vectors are positive.  The 12 regions are disjoint and fill
the planes completely, as shown in Fig.~\ref{fig1}.  On the border
lines between the regions the condition of linear independence between
any three $u$ vectors and any three $v$ vectors is violated.

To summarize, we have learned how to test whether a set of 5 product
vectors $w_i=u_i\otimes v_i$, which is generic in the sense that any
three $u$ vectors are linearly independent and any three $v$ vectors
are also linearly independent, span the kernel of a rank $(4,4)$ PPT
state.  We transform to the standard form defined in
eq.~(\ref{equvnonorth}), by a product transformation and
normalization.  Then the necessary and sufficient condition is that
the parameters $p,q,r,s$ are all real, and that the parameter pairs
$(p,q)$ and $(r,s)$ lie in corresponding regions in the parameter
planes, as shown in Figure~\ref{fig1}.

It should be stressed that these conclusions are based in part on
numerical evidence, and we have no analytical proof which is complete
in every detail.

We will discuss next how to reconstruct a PPT state from the product
vectors in its kernel.

\subsection{Matrix representation relative to a non-orthonormal
  basis}

Consider a basis $\{e_i\}$ consisting of vectors that need not be
orthonormal.  The scalar products
\be{eq089}
g_{ij}=e_i^{\,\dag}e_j
\ee
define the metric tensor $g$ as a Hermitean matrix.  In the usual way,
we write the inverse matrix $g^{-1}$ with upper indices, so that
\be{eq090}
\sum_j g^{ij}g_{jk}=\gd^i_k\;.
\ee
We define the dual vectors $e^j$ such that
\be{eq091}
e^i=\sum_jg^{ji}e_j\;,\qquad
(e^i)^{\dag}=\sum_jg^{ij}e_j^{\,\dag}\;.
\ee
They satisfy the orthogonality relations
\be{eq092}
(e^i)^{\dag}e_j=
e_j^{\,\dag}e^i=
\gd^i_j\;,
\ee
and the completeness relation
\be{eq093}
\iden
=\sum_{i,j} e_ig^{ij}e_j^{\,\dag}
=\sum_j e^je_j^{\,\dag}
=\sum_i e_i(e^i)^{\dag}\;.
\ee
Using the dual basis vectors and the completeness relation we may
write any matrix $A$ as
\be{eq094}
A=\sum_{i,j}e^i\widetilde{A}_{ij}(e^j)^{\dag}
\qquad\mbox{with}\qquad
\widetilde{A}_{ij}=e_i^{\,\dag}Ae_j\;.
\ee

\subsection{Conditions on $\gr$ from product vectors in $\Ker\gr$}

It is possible to construct the rank $(4,4)$ PPT state $\gr$ directly
from 5 product vectors in $\Ker\gr$ without transforming first the
product vectors to the orthogonal form.  We now describe this
construction.

Given three product vectors $w_i=u_i\otimes v_i$ in $\Ker\gr$, with
the restriction that all three $u_i$ and all three $v_i$ are linearly
independent.  Then we have the following product basis of
$\mathbbm{C}^9$, not necessarily orthonormal,
\be{eq120}
e_{ij}=u_i\otimes v_j\;,\qquad
ij=11,12,13,21,22,23,31,32,33\;.
\ee
With respect to this basis we may define matrix elements of $\gr$ like
in eq.~(\ref{eq094}),
\be{eq120a}
\widetilde{\gr}_{ij;kl}=e_{ij}^{\,\dag}\,\gr\,e_{kl}\;.
\ee

In order to count the independent constraints, it is convenient use
the standard form of the product vectors defined in
eq.~(\ref{equvnonorth}).  Now all the constraints from
eq.~(\ref{eq084}) imply that
\be{eq121}
\widetilde{\gr}=\pmatrix{
0 & 0     & 0     & 0     & 0 & 0   & 0     & 0   & 0\cr
0 & a_1   & b_1   & 0     & 0 & 0   & 0     & b_2 & 0\cr
0 & b_1^* & a_2   & 0     & 0 & b_3 & 0     & 0   & 0\cr
0 & 0     & 0     & a_3   & 0 & b_4 & b_5   & 0   & 0\cr
0 & 0     & 0     & 0     & 0 & 0   & 0     & 0   & 0\cr
0 & 0     & b_3^* & b_4^* & 0 & a_4 & 0     & 0   & 0\cr
0 & 0     & 0     & b_5^* & 0 & 0   & a_5   & b_6 & 0\cr
0 & b_2^* & 0     & 0     & 0 & 0   & b_6^* & a_6 & 0\cr
0 & 0     & 0     & 0     & 0 & 0   & 0     & 0   & 0}\;,
\ee
with real diagonal elements $a_1,a_2,\dots,a_6$ and complex
off-diagonal elements $b_1,b_2,\dots,b_6$.  This Hermitean $9\times 9$
matrix contains 18 real parameters, which means that there are
altogether $81-18=63$ independent real constraints.

Including the fourth product vector from eq.~(\ref{equvnonorth}) gives
additional constraints $\widetilde{\gr}w_4=0$, or explicitly
written out,
\be{eq122}
a_1  +b_1+b_2&\!\!\!=&\!\!\!0\;,\nonumber\\
b_1^*+a_2+b_3&\!\!\!=&\!\!\!0\;,\nonumber\\
a_3  +b_4+b_5&\!\!\!=&\!\!\!0\;,\nonumber\\
b_3  +b_4+a_4&\!\!\!=&\!\!\!0\;,\\
b_5^*+a_5+b_6&\!\!\!=&\!\!\!0\;,\nonumber\\
b_2  +b_6+a_6&\!\!\!=&\!\!\!0\;.\nonumber
\ee
Here we have simplified slightly by complex conjugating the 4th and
the 6th equation.  These are complex equations, to be split into real
and imaginary parts.  The real parts are 6 independent equations,
whereas the complex parts are only 5 independent equations.  However,
we get another independent equation as the imaginary part of, for
example, the complex equation
\be{eq123}
(u_1\otimes v_4)^{\dag}\,\gr\,(u_4\otimes v_2)
=a_1+b_1^*+b_2=0\;.
\ee
The end result is that all the off-diagonal matrix elements $b_i$ have
to be real.  Altogether, we get 12 independent real constraints, 6
from the real parts and 6 from the imaginary parts of the equations.

Thus, including the 4th product vector in $\Ker\gr$ increases the
number of independent real constraints from 63 to 75, and reduces the
number of real parameters in $\gr$ from 18 to 6.

The generic case with 5 product vectors is that there are 81
independent constraints, leaving only the trivial solution $\gr=0$.
In order to end up with one possible solution for $\gr$ we have to
choose the parameters $p,q,r,s$ to be real.

When we choose real values for $p,q,r,s$, there is always
(generically) exactly one solution for $\gr$, that is, there are 80
independent constraints.  The problem is that this uniquely determined
matrix $\gr$, or its partial transpose, has in general both positive
and negative eigenvalues.

The condition to ensure that both $\gr\geq 0$ and $\gr^P\geq 0$ (with
the proper choice of sign for $\gr$), when the parameters $p,q,r,s$
are real, is that the pair $(p,q)$ and the pair $(r,s)$ must lie in
corresponding parameter regions, as shown in Fig.~\ref{fig1}.

\subsection{Separable states of rank $(4,4)$}

A separable state of rank 4 has the form
\be{eq124}
\gr=\sum_{i=1}^4\gl_i\,\psi_i\psi_i^{\,\dag}\;,
\ee
with $\gl_i>0$, $\sum_{i=1}^4\gl_i=1$, $\psi_i^{\,\dag}\psi_i=1$, and
$\psi_i=C_i\,\phi_i\otimes\chi_i$ with $C_i$ as a normalization
constant.  In the generic case when any three vectors $\phi_i$ and any
three $\chi_i$ are linearly independent, we may perform an
$\mbox{SL}\otimes\mbox{SL}$ transformation and obtain the standard
form
\be{eq125}
\phi=\pmatrix{
1 & 0 & 0 & 1\cr
0 & 1 & 0 & 1\cr
0 & 0 & 1 & 1},
\qquad
\chi=\pmatrix{
1 & 0 & 0 & 1\cr
0 & 1 & 0 & 1\cr
0 & 0 & 1 & 1}.
\ee
In this standard form the $\chi$ vectors are real, and hence
$\gr^P=\gr$.

The kernel $\Ker\gr$ consists of the vectors that are orthogonal to
all 4 product vectors $\psi_i$, and it contains exactly 6 product
vectors $w_i=N_i\,u_i\otimes v_i$, as follows,
\be{126}
u=\left(\!\begin{array}{cccccc}
 0 & 0 & 0 & 1 & 1 & 1\\
 0 & 1 &-1 & 0 & 0 &-1\\
 1 & 0 & 1 & 0 &-1 & 0
\end{array}\!\right),
\qquad
v=\left(\!\begin{array}{cccccc}
 1 & 1 & 1 & 0 & 0 & 0\\
-1 & 0 & 0 & 1 & 1 & 0\\
 0 &-1 & 0 &-1 & 0 & 1
\end{array}\!\right).
\ee
Note that the product vectors in the kernel of a separable rank
$(4,4)$ PPT state are not generic, in that there are subsets of three
linearly dependent vectors both among the $u$ vectors and among the
$v$ vectors.

The surface of separable states of rank 4 has dimension $35=32+3$,
where 32 is the number of parameters of the group
$\mbox{SL}(3,\mathbbm{C})\otimes\mbox{SL}(3,\mathbbm{C})$ and 3 is the
number of independent coefficients $\gl_i$ in eq.~(\ref{eq124}).

\section{Rank $(5,5)$ PPT states in $3\times 3$ dimensions}
\label{secrank55ppt}

\subsection{The surface of rank $(5,5)$ PPT states}

Since we believe that we understand completely the rank $(4,4)$
entangled states in dimension $3\times 3$, a natural next step is to
try to understand the $(5,5)$ states in the same dimension.

As discussed in the previous section, a generic 5 dimensional subspace
in $3\times 3$ dimensions contains exactly 6 product vectors, which
can be transformed by $\mbox{SL}\otimes\mbox{SL}$ transformations, as
in eqs.~(\ref{eq00mc}) and (\ref{eq00md}), into the standard form
given in eqs.~(\ref{equvnonorth}) and (\ref{equvvecI}), with
$\mbox{SL}\otimes\mbox{SL}$ invariant complex parameters $p,q,r,s$.
Thus, each such subspace belongs to an equivalence class under
$\mbox{SL}\otimes\mbox{SL}$ transformations, and the equivalence
classes are parametrized by 8 real parameters.  There is a discrete
ambiguity in the parametrization, since it depends on the ordering of
the 6 product vectors.

In one given generic 5 dimensional subspace we may construct a 5
dimensional set of rank $(5,5)$ separable states as convex
combinations of the 6 product vectors in the subspace.  However, we
find numerically that the dimension of the surface of rank $(5,5)$ PPT
states with the given subspace as image is not 5 but 8.  We compute
this dimension in the following way.

We search numerically for one rank $(5,5)$ state $\gr$, for example by
the methods described in~\cite{Sollid10a}.  The state we find will
typically be entangled and extremal in $\cP$.  From this state $\gr$
we compute the projections $P,Q,\widetilde{P},\widetilde{Q}$ as
described in Sections~\ref{secbasiclinalg} and~\ref{secrestrpert}, and
we look for perturbations $\gr'=\gr+\ge A$, with $\Tr A=0$, where $A$
satifies both equations
${\rm\bf P}A=PAP=A$ and
$\widetilde{\rm\bf Q}A=(\widetilde{Q}A^P\widetilde{Q})^P=0$,
or equivalently eq.~(\ref{eq013}),
\be{eq130}
({\rm\bf I}-{\rm\bf P}+\widetilde{\rm\bf Q})A=0\;.
\ee
The number of linearly independent solutions for $A$ is the dimension
of the surface of rank $(5,5)$ PPT states at the point $\gr$.

We believe that the dimension 8 can be understood as follows.  We may
fix both subspaces $\Img\gr$ and $\Img\gr^P$, this means that we fix
the projections $P$ and $\widetilde{P}$ and determine $\gr$ as a
solution of the equation
\be{eq130a}
(2{\rm\bf I}-{\rm\bf P}-\widetilde{\rm\bf P})\gr=0\;,
\ee
with $\Tr\gr=1$.  Then there is typically no solution at all for
$\gr$, solutions exist only for special pairs of subspaces.  If now
the two 5 dimensional subspaces are chosen in such a way that a
solution exists, then the solution is (typically) unique, and the
uniqueness means that $\gr$ is an extremal point of $\cP$.

We may fix instead $\Img\gr$ but not $\Img\gr^P$, only the rank of
$\gr^P$.  Then there is a set of solutions for $\gr$ described by 8
real parameters.  It is a natural guess that the role of these 8
parameters is to specify the $\mbox{SL}\otimes\mbox{SL}$ equivalence
class to which the 5 dimensional subspace $\Img\gr^P$ belongs.

In fact, when we fix $\Img\gr$ there is no degree of freedom left
corresponding to $\mbox{SL}\otimes\mbox{SL}$ transformations.  This is
so because the set of product vectors in $\Img\gr$ is discrete and can
not be transformed continuously within the fixed subspace $\Img\gr$.
Hence, the only way to vary the subspace $\Img\gr^P$ without varying
$\Img\gr$ is to vary the equivalence class of $\Img\gr^P$.

\begin{figure}[htp!]
\begin{center}
\includegraphics[width=11cm]{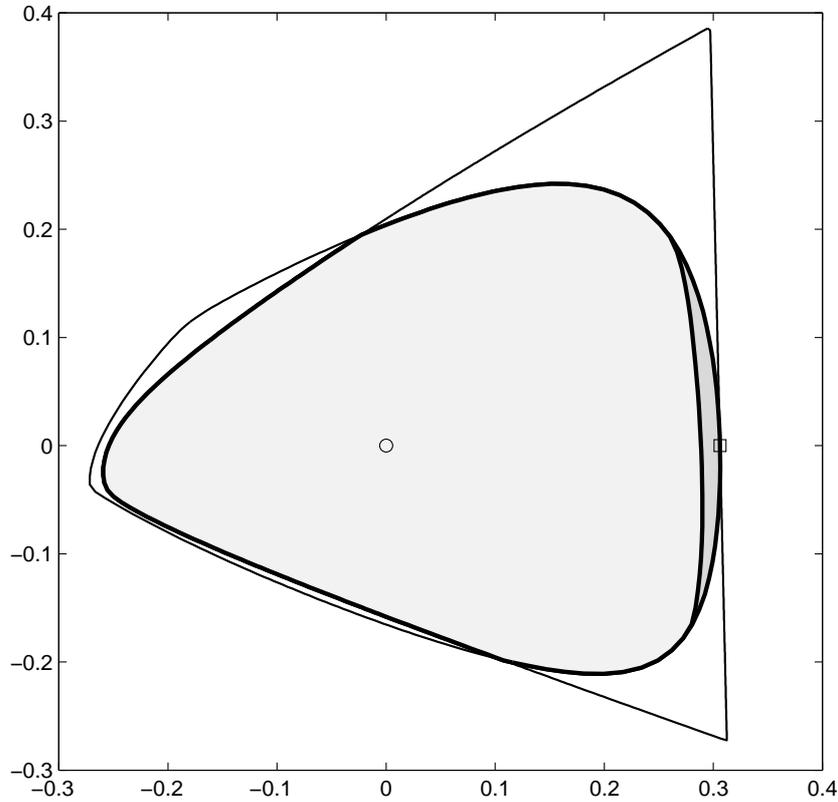}
\end{center}
\caption{\small
Two dimensional section through $\cD$, the set of
density matrices in $3\times 3$ dimensions.
The boundaries of $\cS$, the set of separable states,
and of $\cP$, the set of PPT states,
are both drawn as thick lines.
The boundaries of $\cD$
and of $\cD^P$, the set of
partially transposed density matrices, cross at
two points, they are drawn as thin lines where
they do not coincide with the boundary of $\cP$.
The origin, marked by a small circle, is
the maximally mixed state.
The point marked by a small square is an extremal
rank $(5,5)$ PPT state.
The boundary of $\cD$ is drawn through this point as
a thin straight line.
The boundary of $\cD^P$
is drawn thick at this point, beacause it is also
the boundary of $\cP$.
The lightly shaded region around the origin is $\cS$.
The small and more darkly shaded region close to
the $(5,5)$ state is the difference
between $\cP$ and $\cS$.
Away from this small region,
the boundaries of $\cP$ and $\cS$ are indistinguishable
in the plot.
\label{snittfig1}}
\end{figure}

Figure~\ref{snittfig1} shows a two dimensional section through the set
of density matrices.  The section is defined by the maximally mixed
state, by a randomly selected rank $(5,5)$ entangled and extremal PPT
state $\gr$, and by a direction $A$ through $\gr$ such that the
perturbed state $\gr'=\gr+\ge A$ is a rank $(5,5)$ PPT state for
infinitesimal positive and negative $\ge$, and has $\Img\gr'=\Img\gr$
even for finite $\ge$.  The figure illustrates the fact that the
difference between the sets $\cP$ and $\cS$ is small.  It also
illustrates that the difference between $\cP$ and $\cS$ is largest
close to extremal entangled PPT states.


If we want to allow both $\Img\gr$ and $\Img\gr^P$ to vary, but still
require the ranks of $\gr$ and $\gr^P$ to be 5, the equation to be
solved for the perturbation $A$ is
\be{eq131}
({\rm\bf Q}+\widetilde{\rm\bf Q})A=0\;.
\ee
In this case the number of linearly independent solutions for $A$ is
48, and this is the dimension of the surface of all rank $(5,5)$ PPT
states.

We understand the dimension 48 as follows.  There are $8+8=16$
parameters for the $\mbox{SL}\otimes\mbox{SL}$ equivalence classes of
the subspaces $\Img\gr$ and $\Img\gr^P$.  And there are 32~parameters
for the $\mbox{SL}(3,\mathbbm{C})\otimes\mbox{SL}(3,\mathbbm{C})$
transformations.

To summarize, an extremal (and hence entangled) rank $(5,5)$ PPT state
$\gr$ is uniquely determined by eq.~(\ref{eq130a}), as soon as we
specify the 5 dimensional subspaces $\Img\gr$ and $\Img\gr^P$.  Each
subspace $\Img\gr$ and $\Img\gr^P$ is determined by 8 real
$\mbox{SL}\otimes\mbox{SL}$ invariant parameters and an
$\mbox{SL}\otimes\mbox{SL}$ transformation.

According to our understanding, based on numerical studies, which is
so far only a plausible hypothesis, the 8 invariant parameters can be
chosen independently for $\Img\gr$ and $\Img\gr^P$, but the
$\mbox{SL}\otimes\mbox{SL}$ transformations can not be chosen
independently.  However, we do not know the relation that clearly
exists between $\Img\gr$ and $\Img\gr^P$.

In other words, we do not know any explicit procedure for constructing
the most general rank $(5,5)$ PPT states.  Therefore we turn next to a
more restricted problem.

\subsection{Perturbing from rank $(4,4)$ to rank $(5,5)$}
\label{subsecperturbing4455}

We will see now how to construct $(5,5)$ states that are
infinitesimally close to $(4,4)$ states.

Consider once more an infinitesimal perturbation
$\gr'=\gr+\ge A$, this time with $\gr$ as the rank $(4,4)$ state
defined in eq.~(\ref{eq085}), involving the standard real product
vectors defined in eq.~(\ref{equvorth}).  The most general case is
equivalent to this special case by some $\mbox{SL}\otimes\mbox{SL}$
transformation.

An extra bonus of this special choice of $\gr$ is that $\gr^P=\gr$.
In the notation we have used above, we have projections
$P=\widetilde{P}$ on $\Img\gr=\Img\gr^P$ and $Q=\widetilde{Q}$ on
$\Ker\gr=\Ker\gr^P$.

\subsubsection*{Conditions on the perturbation matrix $A$}

By eq.~(\ref{eq00j}), the condition for $\gr'$ to have rank 5 is that
\be{eq140}
QAQ=\ga\,ww^{\dag}\;,
\ee
where $\ga$ is a real number, $\ga\neq 0$, and
\be{eq141}
w=\sum_{i=1}^5 c_iw_i\;,
\ee
with complex coefficients $c_i$ such that
\be{eq142}
w^{\dag}w=\sum_{i=1}^5 |c_i|^2=1\;.
\ee
Similarly, the condition for $(\gr')^P$ to have rank 5 is that
\be{eq143}
\widetilde{Q}A^P\widetilde{Q}=QA^PQ=\gb\,zz^{\dag}\;,
\ee
where $\gb$ is real, $\gb\neq 0$, and
\be{eq144}
z=\sum_{i=1}^5 d_iw_i
\qquad\mbox{with}\qquad
z^{\dag}z=\sum_{i=1}^5 |d_i|^2=1\;.
\ee
Note that the possibilities that $\gr'$ has rank either $(4,5)$ or
$(5,4)$ are included if we allow either $\ga$ or $\gb$ to be zero.

By eq.~(\ref{eq00kd}), there is one extra product vector in
$\Ker\gr=\Ker\gr^P$, it may be written as
\be{eqw6}
w_6=\sum_{i=1}^5 a_iw_i\;,
\ee
this time with real coefficients $a_i$.  Since $w_i=N_iu_i\otimes v_i$
with $N_i$ real and $v_i$ real for $i=1,2,\ldots,6$, we have for any
Hermitean matrix $A$ that
\be{eq145}
w_i^{\,\dag}Aw_i=w_i^{\,\dag}A^Pw_i\;.
\ee
By the definition of the projection $Q$ we have that $Qw_i=w_i$ for
$i=1,2,\ldots,6$.  It follows then from eq.~(\ref{eq140}) that
\be{eq146}
w_i^{\,\dag}Aw_i=
w_i^{\,\dag}QAQw_i=
\ga\,|w_i^{\,\dag}w|^2\;,
\ee
and from eq.~(\ref{eq143}) that
\be{eq147}
w_i^{\,\dag}A^Pw_i=
w_i^{\,\dag}QA^PQw_i=
\gb\,|w_i^{\,\dag}z|^2\;.
\ee
Together with eq.~(\ref{eq145}) this gives the equations
\be{eq148}
\ga\,|c_i|^2=\gb\,|d_i|^2
\ee
for $i=1,2,\ldots=5$, and the 6th equation
\be{eq149}
\ga\left|\sum_{i=1}^5 a_ic_i\right|^2=
\gb\left|\sum_{i=1}^5 a_id_i\right|^2\;.
\ee
It follows further that
\be{eq150}
\ga
=\sum_{i=1}^5 \ga\,|c_i|^2
=\sum_{i=1}^5 \gb\,|d_i|^2
=\gb\;,
\ee
and that
\be{eq151}
|c_i|=|d_i|
\qquad\mbox{for}\qquad
i=1,2,\ldots,5\;.
\ee
Thus, the coefficient $d_i$ can differ from $c_i$ only by a phase
factor.  The total of 5 phase factors are reduced to 4 independent
phase factors by the extra equation
\be{eq152}
\left|\sum_{i=1}^5 a_ic_i\right|=
\left|\sum_{i=1}^5 a_id_i\right|\;.
\ee

For infinitesimal values of $\ge$, both $\gr'$ and $(\gr')^P$ will
have four eigenvalues infinitesimally close to $1/4$ and one
eigenvalue close to zero, which is $\ge\ga$ for $\gr'$ and $\ge\gb$
for $(\gr')^P$.  This eigenvalue is the same for $\gr'$ and
$(\gr')^P$, since $\ga=\gb$.  With $\ga>0$ this means that both
$\gr'\geq 0$ and $(\gr')^P\geq 0$ for $\ge>0$, but not for $\ge<0$.
Thus, we get automatically a PPT state of rank $(5,5)$, we never get
rank $(5,4)$ or $(4,5)$.  Also it never happens that $\gr'$ is not a
PPT state for the reason that one of $\gr'$ or $(\gr')^P$ has a
negative eigenvalue.

For a more general rank $(4,4)$ state $\gr$, which is obtained by some
$\mbox{SL}\otimes\mbox{SL}$ transformation from a state of the special
type discussed here, the smallest positive eigenvalues of $\gr'$ and
$(\gr')^P$ are no longer equal.  But they are still tied together in
such a way that they go to zero simultaneously when we move along the
surface of $(5,5)$ states and approach its boundary.  The boundary
must therefore consist of $(4,4)$ states.

\subsubsection*{Computing $A$}

Define $W=ww^{\dag}$ and $Z=zz^{\dag}$, in the same notation as above.
These are both projections, $W^2=W$ and $Z^2=Z$, with $QW=WQ=W$ and
$QZ=ZQ=Z$.  It follows from eq.~(\ref{eq140}) and
eq.~(\ref{eq143}) that
\be{eq153}
WAW&\!\!\!=&\!\!\!WQAQW=\ga W^3=\ga W=QAQ\;,\nonumber\\
ZA^PZ&\!\!\!=&\!\!\!ZQA^PQZ=\gb Z^3=\gb Z=QA^PQ\;.
\ee
Like in eq.~(\ref{eq002}) and eq.~(\ref{eq009}) we define
\be{eq154}
{\rm\bf P}X=PXP\;,\quad
{\rm\bf Q}X=QXQ\;,\quad
\widetilde{\rm\bf P}X=(PX^PP)^P\;,\quad
\widetilde{\rm\bf Q}X=(QX^PQ)^P\;,
\ee
and furthermore,
\be{eq154a}
{\rm\bf W}X=WXW\;,\qquad
\widetilde{\rm\bf Z}X=(ZX^PZ)^P\;.
\ee
We may also define ${\rm\bf S}={\rm\bf Q}-{\rm\bf W}$ and
$\widetilde{\rm\bf S}=\widetilde{\rm\bf Q}-\widetilde{\rm\bf Z}$,
these are again orthogonal projections on $H_N$.

The least restrictive conditions we may impose on $A$ are now that
both ${\rm\bf S}A=0$ and $\widetilde{\rm\bf S}A=0$, or equivalently,
\be{eq155}
({\rm\bf S}+\widetilde{\rm\bf S})A=0\;.
\ee
To compute $A$ from this equation we introduce an orthonormal basis in
the real Hilbert space $H_N$.  Relative to this basis, the operator
${\rm\bf S}+\widetilde{\rm\bf S}$ is represented by a real symmetric
positive semidefinite matrix, which has a complete set of real
eigenvectors with real eigenvalues.  We choose $A$ as an eigenvector
of ${\rm\bf S}+\widetilde{\rm\bf S}$ with eigenvalue zero.

Apart from the trivial solution $A=\gr$, we find 37 linearly
independent solutions of eq.~(\ref{eq155}).  36 out of these 37 are
perturbations that give $\gr'=\gr+\ge A$ as a rank $(4,4)$ state both
for $\ge>0$ and $\ge<0$.  They do not depend on either vector $w$ or
$z$, since they satisfy the conditions ${\rm\bf Q}A=0$ and
$\widetilde{\rm\bf Q}A=0$.  But because
${\rm\bf W}={\rm\bf W}{\rm\bf Q}$ and
$\widetilde{\rm\bf Z}=\widetilde{\rm\bf Z}\widetilde{\rm\bf Q}$ they
also satisfy the conditions
\be{eq155a}
{\rm\bf W}A={\rm\bf W}{\rm\bf Q}A=0\;,\qquad
\widetilde{\rm\bf Z}A=\widetilde{\rm\bf Z}\widetilde{\rm\bf Q}A=0\;,
\ee
and hence eq.~(\ref{eq155}).  The number 36 is the dimension of the
surface of rank $(4,4)$ extremal PPT states, as noted in
Subsection~\ref{subsecUPB}.  The 37th independent solution is the one
giving a rank $(5,5)$ extremal PPT state.

A more restricted class of perturbations consists of those where we
fix the 5 dimensional subspace $\Img\gr'$ to be the direct sum of the
4 dimensional subspace $\Img\gr$ and the one dimensional subspace of
the vector $w$.  The projection on $\Img\gr'$ is then
\be{eq156}
P_5=P+W\;,
\ee
and the partial condition on $A$ is that ${\rm\bf P}_5A=A$, when we
define
\be{eq157}
{\rm\bf P}_5X=P_5XP_5\;.
\ee
The full condition on $A$ is that
\be{eq158}
({\rm\bf P}_5-\widetilde{\rm\bf S})A=
A\;.
\ee

Again apart from the trivial solution $A=\gr$, we find 5 linearly
independent solutions of eq.~(\ref{eq158}), of which 4 give $\gr'$ as
a rank $(4,4)$ state both for $\ge>0$ and $\ge<0$.  The 5th
independent solution is the one giving $\gr'$ as a rank $(5,5)$
extremal PPT state.

The 4 directions that only give new $(4,4)$ states are easily
identified, since they do not depend on the vector $z$.  To find them
we simply repeat the calculation with a ``wrong'' $z$, violating the
conditions~(\ref{eq151}) and~(\ref{eq152}).  In this way we find no
$(5,5)$ state, but we find the same set of perturbations into $(4,4)$
states.  The number 4 is the dimension of the surface of $(4,4)$
states with image within the fixed 5 dimensional subspace projected
out by the projector $P_5$.

There is a natural explanation of why this surface has dimension 4.
In fact, when we fix $P_5$
%
%
and look for $(4,4)$ states with image within this fixed 5 dimensional
subspace, we eliminate all degrees of freedom corresponding to
$\mbox{SL}\otimes\mbox{SL}$ transformations.  But we still allow
variations of the 4 real $\mbox{SL}\otimes\mbox{SL}$ invariant
parameters that are needed to define a rank $(4,4)$ state.

We conclude that for fixed vectors $w$ and $z$ there is one direction
away from the surface of rank $(4,4)$ extremal PPT states and into the
surface of rank $(5,5)$ extremal PPT states.

For a fixed vector $w$ there is a 4 parameter family of acceptable
vectors $z$.  Recall that these 4 parameters determine the 5 relative
phases between the coefficients $c_i$ in eq.~(\ref{eq141}) and the
corresponding coefficients $d_i$ in eq.~(\ref{eq144}).

The vector $w$ is an arbitrary vector in the 5 dimensional kernel of
the unperturbed state $\gr$, hence it contains 4 complex parameters,
or 8 real parameters, after we take out an uninteresting complex
normalization factor.  Altogether, there are $8+4=12$ independent
directions away from the 36 dimensional surface of rank $(4,4)$ PPT
states and into the surface of rank $(5,5)$ PPT states.

When we perturb an arbitrary rank $(5,5)$ PPT state in such a way that
we preserve the ranks of the state and its partial transpose, we find
numerically that the surface of rank $(5,5)$ PPT states has dimension
48.  The fact that $48=36+12$ is consistent with the hypothesis that
we can reach every rank $(5,5)$ PPT state if we start from a rank
$(4,4)$ PPT state and move continuously along the surface of rank
$(5,5)$ PPT states.

\section{Rank $(6,6)$ entangled PPT states in $4\times 4$ dimensions}
\label{secrank66ppt}

We will discuss in some detail one more example of the relation
between PPT states and product vectors.  According to
eq.~(\ref{eq00kc}), the rank $(6,6)$ PPT states in $4\times 4$
dimensions represent just the limiting case with a finite number of
product vectors in the kernel, in this respect they are similar to the
rank $(4,4)$ states in $3\times 3$ dimensions.

The kernel of a rank 6 state in 16 dimensions has dimension 10, and
the generic case, according to eq.~(\ref{eq00kd}), is that it contains
exactly 20 product vectors, any 10 of which are linearly independent.
We will see here that the product vectors in the kernel put such
strong restrictions on the state that the rank $(6,6)$ PPT state may
be reconstructed uniquely from only 7 product vectors in its kernel.

To see how it works, take a set of product vectors in $4\times 4$
dimensions.  We may take random product vectors, or else a set of
product vectors with the special property that they belong to
$\Ker\gr$ where $\gr$ is a rank $(6,6)$ PPT state.  We find
numerically that the number of constraints generated by fewer than 7
product vectors is the same in both cases.  We find the following
numbers.

From 4 product vectors assumed to lie in $\Ker\gr$ for an unknown
$\gr$, or actually lying in $\Ker\gr$ for a known $\gr$, we get 172
independent constraints on $\gr$ of the form given in
eq.~(\ref{eq084}).  These constraints leave 84 free real parameters in
$\gr$, before we normalize and set $\Tr\gr=1$.

From 5 product vectors in $\Ker\gr$ we get 205 independent
constraints, leaving 51 parameters in $\gr$.

From 6 product vectors in $\Ker\gr$ we get 234 independent
constraints, and 22 parameters in $\gr$.

Finally, 7 product vectors in $\Ker\gr$ give either 255 or 256
independent constraints, and either 1 or 0 real parameters in $\gr$.
If there is one parameter left, it is a proportionality constant, to
be fixed by the normalization condition $\Tr\gr=1$.


The standard form of 7 product vectors in $4\times 4$ dimensions,
generalizing eq.~(\ref{equvnonorth}), is the following,
\be{equvnonorth4x4}
u=\pmatrix{
1 & 0 & 0 & 0 & 1 & 1   & 1\cr
0 & 1 & 0 & 0 & 1 & p_1 & p_4\cr
0 & 0 & 1 & 0 & 1 & p_2 & p_5\cr
0 & 0 & 0 & 1 & 1 & p_3 & p_6},
\qquad
v=\pmatrix{
1 & 0 & 0 & 0 & 1 & 1   & 1\cr
0 & 1 & 0 & 0 & 1 & p_7 & p_{10}\cr
0 & 0 & 1 & 0 & 1 & p_8 & p_{11}\cr
0 & 0 & 0 & 1 & 1 & p_9 & p_{12}}.
\ee
There are 12 complex parameters $p_1,p_2,\ldots,p_{12}$, that is, 24
real parameters.  These are invariant in the sense that we can not
change them by
$\mbox{SL}(4,\mathbbm{C})\otimes\mbox{SL}(4,\mathbbm{C})$
transformations.

Not just any arbitrary set of 7 product vectors defines a rank $(6,6)$
PPT state.  We arrive at this conclusion not only because we find
numerically that 7 generic product vectors allow only $\gr=0$ as
solution of all the constraint equations, but also because a dimension
counting shows that we need less than 24 invariant parameters in order
to parametrize the rank $(6,6)$ PPT states.

Take one known rank $(6,6)$ PPT state $\gr$ and perturb it into
another rank $(6,6)$ PPT state $\gr'=\gr+\epsilon A$ with $\ge$
infinitesimal.  Here $A$ must be a solution of eq.~(\ref{eq012}), with
operators ${\rm\bf Q}$ and $\wt{\rm\bf Q}$ defined relative to $\gr$
as explained.  The number of linearly independent solutions for $A$,
found numerically, is 76, including the trivial solution $A=\gr$.
This shows that the surface of rank $(6,6)$ PPT states has 75 real
dimensions.

Of these 75 dimensions, 60 dimensions result from product
transformations $\gr\mapsto V\gr V^{\dag}$ with
$V=V_A\otimes V_B$ and $V_A,V_B\in\mbox{SL}(4,\mathbbm{C})$.  The
remaining 15 dimensions must correspond to 15
$\mbox{SL}\otimes\mbox{SL}$ invariant parameters of the 7 product
vectors.

It is also worth noting that, by the counting explained in the next
section, the set of 6 dimensional subspaces of $\mathbbm{C}^{16}$ has
real dimension $16^2-6^2-10^2=120$, much larger than the dimension 75
of the surface of $(6,6)$ states.  Thus, not every 6 dimensional
subspace of $\mathbbm{C}^{16}$ is the host of a rank $(6,6)$ PPT
state, as one would expect from the analogy to the case of the rank
$(4,4)$ PPT states in $\mathbbm{C}^9$.

\section{Dimension counting}
\label{secdimcount}

We will describe in this section how to compute numerically the
dimensions of surfaces of PPT states of given ranks.  We list some
numerical results, and discuss how they may be understood in most
cases by a simple counting of constraints, assuming the constraints to
be independent.

We start with a useful exercise.  We want to compute the real (as
opposed to complex) dimension of the set of all $r$ dimensional
subspaces of an $N$ dimensional complex Hilbert space.

First note that the unitary group $\mbox{U}(k)$ has $k^2$ real
dimensions.  Take an orthonormal basis of the Hilbert space.  The
first $r$ basis vectors define an $r$ dimensional subspace, the
orthogonal complement of which is defined by the last $s=N-r$ basis
vectors.  A $\mbox{U}(N)$ transformation transforms this basis into
another orthonormal basis, but the $\mbox{U}(r)$ transformations
within the first $r$ basis vectors, and the $\mbox{U}(s)$
transfomations within the last $s$ basis vectors, do not change either
subspace.  It follows that the dimension of the set of $r$ dimensional
subspaces, equal to the dimension of the set of $s$ dimensional
subspaces, is
\be{eq801}
d=N^2-r^2-s^2=2rs\;.
\ee

Assuming that we have found a PPT state $\gr$ of rank $(m,n)$, it lies
on a surface of rank $(m,n)$ PPT states.  We compute the dimension of
the surface at this point by counting the number of independent
solutions $A$ of eq.~(\ref{eq012}),
\be{eq812}
({\rm\bf Q}+\wt{\rm\bf Q})A=0\;,
\ee
equivalent to the two equations
\be{eq813}
{\rm\bf Q}A=QAQ=0\;,\qquad
\wt{\rm\bf Q}A=(\wt{Q}A^P\wt{Q})^P=0\;.
\ee
We have to throw away the trivial solution $A=\gr$.  We get a lower
bound for the dimension if we assume that the constraints on $A$ from
the two equations in eq.~(\ref{eq813}) are independent.  The equation
$QAQ=0$ represents $(N-m)^2$ real constraints, since $Q$ is the
orthogonal projection on the $N-m$ dimensional subspace $\Ker\gr$.
Similarly, the equation $\wt{Q}A^P\wt{Q}=0$ represents $(N-n)^2$ real
constraints, since $\wt{Q}$ is the orthogonal projection on the $N-n$
dimensional subspace $\Ker\gr^P$.  Because the constraints are not
necessarily independent, we get the following lower bound for the
dimension,
\be{eq814}
d\geq N^2-(N-m)^2-(N-n)^2-1\;.
\ee

Take $N=3\times 3=9$ as an example.  We find numerically that
eq.~(\ref{eq814}) holds with equality for all ranks from the full rank
$(m,n)=(9,9)$ down to $(m,n)=(5,5)$.  In particular, for rank $(5,5)$
the dimension of the surface is
\be{eq815}
d=9^2-4^2-4^2-1=48\;.
\ee
By eq.~(\ref{eq801}) the set of 5 dimensional subspaces has dimension
40, hence we should expect to find an 8 dimensional surface of rank
$(5,5)$ PPT states in every 5 dimensional subspace.  And that is
actually what we find.

For rank $(4,4)$ the constraints are not all independent, and we have
the strict inequality
\be{eq816}
d=36>9^2-5^2-5^2-1=30\;.
\ee
The set of 4 dimensional subspaces has again dimension 40, hence there
can not exist rank $(4,4)$ PPT states in every 4 dimensional subspace.
There are $40-36=4$ constraints restricting the 4 dimensional
subspaces supporting rank $(4,4)$ PPT states, and in each 4
dimensional subspace there can exist at most one unique such state.
The 4 constraints are the conditions that the 4 parameters $a,b,c,d$
in eq.~(\ref{equvorth}), or $p,q,r,s$ in eq.~(\ref{equvnonorth}), have
to be real.

If we want to compute the dimension of the surface of rank $(m,n)$ PPT
states with fixed image space, we have to count the independent
solutions of eq.~(\ref{eq013}),
\be{eq817}
({\rm\bf I}-{\rm\bf P}+\wt{\rm\bf Q})A=0\;,
\ee
equivalent to the two equations
\be{eq818}
{\rm\bf P}A=PAP=A\;,\qquad
\wt{\rm\bf Q}A=(\wt{Q}A^P\wt{Q})^P=0\;.
\ee
The equation ${\rm\bf P}A=A$ leaves $m^2$ real parameters in $A$ and
represents $N^2-m^2$ real constraints, as is visualized in
eq.~(\ref{eq008}).  The lower bound on the dimension is therefore
\be{eq819}
d\geq N^2-(N^2-m^2)-(N-n)^2-1=m^2-(N-n)^2-1\;.
\ee

In the above example with $N=9$ and $(m,n)=(5,5)$ we find numerically
$d=8$, as already mentioned, so that the inequality in
eq.~(\ref{eq819}) holds as an equality.  With $(m,n)=(4,4)$, on the
other hand, we get
\be{eq820}
d=0\geq 4^2-5^2-1=-10\;.
\ee

\section{Numerical integration}
\label{secnumint}

In this section we will describe a numerical method for tracing curves
on a surface of PPT states of fixed ranks $(m,n)$.  This is a tool for
studying the geometry of the surface, for example by tracing geodesics
to see how they curve, or studying how the surface approaches a
boundary consisting of states of lower ranks.

\subsection{Equations of motion}

The perturbation expansion $\gr(t+\ge)=\gr(t)+\ge A$ for $\gr=\gr(t)$
is equivalent to the differential equation
\be{eq901}
\dot{\gr}=A\;.
\ee
We use the notation
\be{eq901a}
\dot{\gr}=\frac{{\rm d}\gr}{{\rm d}t}\;,\qquad
\dot{\gr}^+=\frac{{\rm d}\gr^+}{{\rm d}t}\;.
\ee
We defined the pseudoinverse $\gr^+$ in eq.~(\ref{eq00ba}), in order
to define $P=\gr^+\gr=\gr\gr^+$ and $Q=\iden-P$, the orthogonal
projections on $\Img\gr$ and $\Ker\gr$, respectively.  There are
similar relations for $\widetilde{P}$, the projection on $\Img\gr^P$,
and $\widetilde{Q}=\iden-\widetilde{P}$, the projection on
$\Ker\gr^P$.  We defined orthogonal projections on $H_N$, the space of
Hermitean matrices, in eq.~(\ref{eq002}) and eq.~(\ref{eq009}).

If $X$ is a Hermitean matrix with $\Img X\subset\Img\gr$ then
$X=PX=XP$, or equivalently, $QX=XQ=0$.  Assuming that these relations
hold at any ``time'' $t$ we may differentiate and get that
\be{eq901b}
\dot{X}=\dot{P}X+P\dot{X}=\dot{X}P+X\dot{P}\;.
\ee
Equivalently,
\be{eq901c}
Q\dot{X}=\dot{P}X\;,\qquad
\dot{X}Q=X\dot{P}\;.
\ee
Multiplication by $Q$ from the left and from the right gives that
\be{eq901d}
Q\dot{X}Q=0\;.
\ee
It follows further that
\be{eq901da}
\dot{X}=(P+Q)\dot{X}(P+Q)
=P\dot{X}P+X\dot{P}+\dot{P}X\;.
\ee

The special case $X=\gr$ gives the equation
\be{eq901e}
QAQ=0
\ee
as a consistency condition for eq.~(\ref{eq901}) with the relations
$\gr=P\gr=\gr P$.  Eq.~(\ref{eq901e}) is the same as
eq.~(\ref{eq00ia}), the condition for the rank of $\gr$ to be
constant.  We may want to replace it with the stronger condition that
$\Img\gr$ should be constant, eq.~(\ref{eq00ha}),
\be{eq901ea}
PAP=A\;.
\ee

Setting $X=\gr$ in eq.~(\ref{eq901c}) gives the equations
\be{eq901f}
QA=\dot{P}\gr\;,\qquad
AQ=\gr\dot{P}\;,
\ee
and multiplication by $\gr^+$ gives that
\be{eq901g}
QA\gr^+=\dot{P}P\;,\qquad
\gr^+AQ=P\dot{P}\;.
\ee
Differentiating the equation $P=P^2$ gives that
$\dot{P}=\dot{P}P+P\dot{P}$, hence
\be{eq901h}
\dot{P}=QA\gr^++\gr^+AQ\;.
\ee

Differentiating the relation $\gr^+=\gr^+\gr\gr^+$ we get that
\be{eq901i}
\dot{\gr}^+=\dot{\gr}^+P+\gr^+A\gr^++P\dot{\gr}^+\;.
\ee
When we left and right multiply here by $P$ we obtain the relation
\be{eq901j}
P\dot{\gr}^+P=-\gr^+A\gr^+\;.
\ee
Hence, using eq.~(\ref{eq901da}) with $X=\gr^+$, together with
eq.~(\ref{eq901h}), we get that
\be{eq901k}
\dot{\gr}^+=QA(\gr^+)^2+(\gr^+)^2AQ-\gr^+A\gr^+\;.
\ee

The equations~(\ref{eq901}), (\ref{eq901h}), and (\ref{eq901k}) may be
integrated together, as soon as we specify how to calculate $A$ as a
function of $\gr$.  There are, of course, equations similar to
(\ref{eq901h}) and (\ref{eq901k}) that hold for the projection
$\widetilde{P}$ related to the partial transpose $\gr^P$, and for the
pseudoinverse $(\gr^P)^+$.

As a specific example, consider how to generate a curve $\gr=\gr(t)$
lying on the 48 dimensional surface in $H_N$ of rank $(5,5)$ PPT
states in $3\times 3$ dimensions passing through a given state
$\gr(0)$.  We then have to satisfy the two conditions on $A$ that
\be{eq902a}
{\rm\bf Q}A=QAQ=0\;,\qquad
\widetilde{\rm\bf Q}A=(\widetilde{Q}A^P\widetilde{Q})^P=0\;.
\ee
Or equivalently eq.~(\ref{eq012}),
\be{eq903a}
({\rm\bf Q}+\widetilde{\rm\bf Q})A=0\;.
\ee

Alternatively, we may want to generate a curve that follows the 8
dimensional surface in $H_N$ of rank $(5,5)$ PPT states such that the
5 dimensional subspace $\Img\gr$ is kept fixed, but the 5 dimensional
subspace $\Img\gr^P$ is allowed to change.  This means that we replace
the condition $QAQ=0$ by the condition $PAP=A$.  The single condition
to be satisfied is then eq.~(\ref{eq013}),
\be{eq905}
({\rm\bf I}-{\rm\bf P}+\widetilde{\rm\bf Q})A=0\;.
\ee
In this case $P$ is constant but $\widetilde{Q}=\widetilde{Q}(t)$ may
vary as a function of $t$.

\subsection{Geodesic equations}

Both conditions~(\ref{eq903a}) and~(\ref{eq905}) are of the form
\be{eq906}
{\rm\bf T}A=0\;.
\ee
Differentiating this equation gives that
\be{eq907}
{\rm\bf T}\dot{A}+\dot{\rm\bf T}A=0\;.
\ee
It follows that
\be{eq908}
\dot{A}=B-{\rm\bf T}^+\dot{\rm\bf T}A\;,
\ee
where ${\rm\bf T}^+$ is the pseudoinverse of ${\rm\bf T}$, and $B$ is
an arbitrary Hermitean matrix with ${\rm\bf T}B=0$.

By definition, a geodesic on an embedded surface (think of a great
circle on the surface of a sphere as an example) is a curve which does
not change its direction on the surface.  Hence, it changes direction
in the embedding space only as much as it has to in order to stay on
the surface.  This would mean that we choose $B=0$ in
eq.~(\ref{eq908}).  Or if we normalize $A$ to unit length, fixing
$\Tr A^2=1$, we set $B=\ga A$ and choose $\ga$ such that
$\Tr(\dot{A}A)=0$.

\subsection{Numerical results}

We have done numerical integrations by a standard fourth order
Runge--Kutta method.  With $\gr$ and $A$ of order one and time steps
of order $10^{-4}$ this gives a precision of order $10^{-16}$, which
is the machine precision.

\begin{figure}[p!]
\begin{center}
\includegraphics[width=11cm]{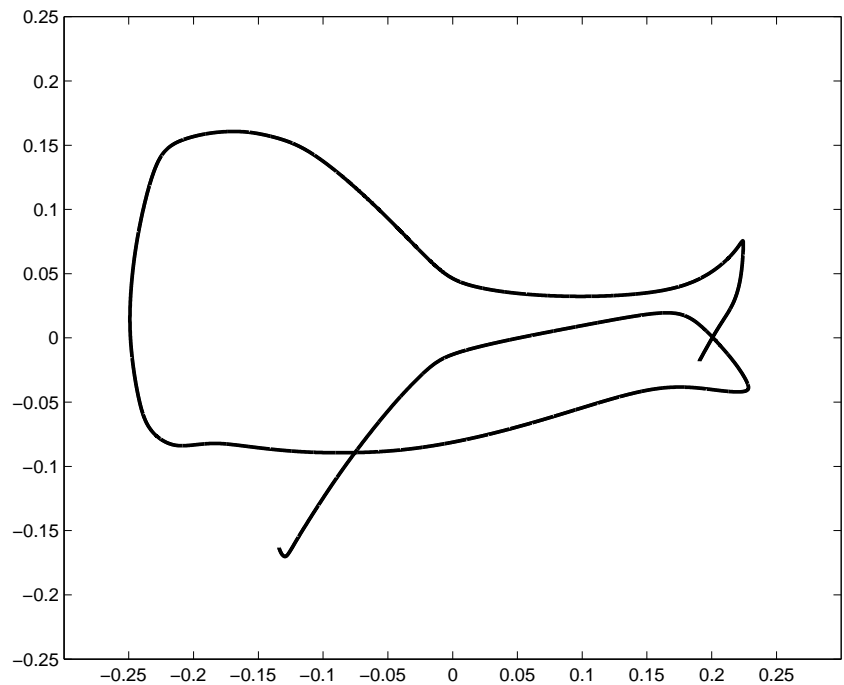}
\end{center}
\caption{\small
A projection of a geodesic curve on the 8 dimensional
surface of rank $(5,5)$ PPT states with a fixed image space.
We have made a principal component analysis and plotted
the two largest principal components.
The curve starts middle right and ends lower left. 
\label{blaahval}}
\end{figure}

\begin{figure}[p!]
\begin{center}
\includegraphics[width=9cm]{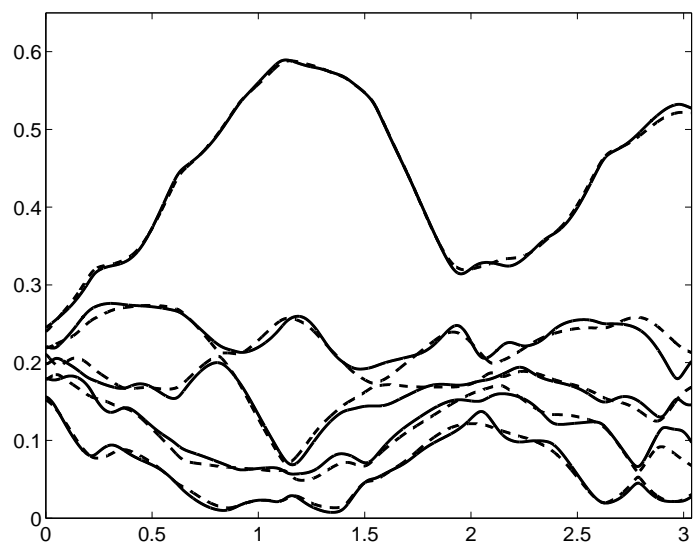}
\end{center}
\caption{\small
Variation along the curve in Fig.~\ref{blaahval}
of the 5 nonzero eigenvalues of the density matrix
(full drawn lines) and its partial transpose (broken lines).
The abscissa is the arc length along the curve.
\label{blaahvalev}}
\end{figure}

Figure~\ref{blaahval} shows a geodesic curve $\gr(t)$ on the 8
dimensional curved surface of rank $(5,5)$ PPT states with
$\Img\gr(t)$ constant.  Figure~\ref{blaahvalev} shows the 5 nonzero
eigenvalues of $\gr$ and $\gr^P$.  The condition that one eigenvalue
of either $\gr$ or $\gr^P$ goes to zero defines the boundary of the
surface.  We see that the curve approaches the boundary twice, but
turns around each time and continues in the interior.  The eigenvalue
spectra of $\gr$ and $\gr^P$ are remarkably similar, yet they are not
identical.  When both $\gr$ and $\gr^P$ simultaneously get one
dominant eigenvalue, we interpret it as an indication that $\gr$
approaches a pure product state.

It is quite natural that a geodesic chosen at random will not hit the
boundary, since the boundary consists of rank $(4,4)$ PPT states and
has dimension 4, while the surface itself has dimension 8.  In order
to hit the boundary we can not follow a geodesic, we have to integrate
the equation $\dot{\gr}=A$ and choose the direction $A$ in such a way
that the smallest positive eigenvalue of $\gr$ goes to zero.  When we
do so, the smallest eigenvalue of $\gr^P$ goes to zero simultaneously
with the eigenvalue of $\gr$, although the ratio between the two
eigenvalues goes to a value different from one.  Hence the curve ends
at a $(4,4)$ state on the boundary.  The explanation for this coupling
of eigenvalues of $\gr$ and $\gr^P$ was given in
Subsection~\ref{subsecperturbing4455}.

\section{Summary}

The work presented here is part of an ongoing programme to study
quantum entanglement in mixed states.  We have studied here low rank
entangled PPT states using perturbation theory and the close relation
between PPT states and product vectors.

One result obtained is an understanding of how to construct rank
$(5,5)$ PPT states in $3\times 3$ dimensions by perturbing rank
$(4,4)$ states.  We use perturbation theory to study surfaces of PPT
states of given ranks, and in particular to compute dimensions of such
surfaces, for example the surface of $(5,5)$ states.  However, it is
still an unsolved problem how to construct general rank $(5,5)$ PPT
states that are not close to rank $(4,4)$ states.  We are even farther
from a full understanding of higher rank PPT states in $3\times 3$
dimensions, or in higher dimensions.

A special class of PPT states are those of special ranks so that their
kernel is spanned by product vectors and contains a finite number of
product vectors.  We have shown that these states may be reconstructed
uniquely from a subset of the product vectors in the kernel, and the
number of product vectors needed may be smaller than the dimension of
the kernel.  This result raises new interesting questions to be
answered by future research, for example, how to identify finite sets
of product vectors that define PPT states with these product vectors
in their kernel.

\section*{Acknowledgments}

We acknowledge gratefully research grants from The Norwegian
University of Science and Technology (Leif Ove Hansen) and from The
Norwegian Research Council (Per {\O}yvind Sollid).
Leif Ove Hansen, Jan Myrheim, and Per {\O}yvind Sollid also want to
thank NORDITA for hospitality during the workshop and conference on
quantum information in Stockholm in September and October 2010.

\vspace*{-1\baselineskip}

\clearpage

\pagebreak




\begin{thebibliography}{99}


\bibitem{RPMKHorodecki}
R.~Horodecki, P.~Horodecki, M.~Horodecki, and K.~Horodecki,
{\em Quantum entanglement},
Rev.~Mod.~Phys.~{\bf 81}, 865--942 (2009).

\bibitem{Bell64}
J.S.~Bell,
{\em On the Einstein Podolsky Rosen paradox},
Physics~{\bf 1}, 195 (1964).

\bibitem{GHZ89}
D.M.~Greenberger, M.A.~Horne, and A.~Zeilinger,
{\em Going beyond Bell's theorem}.
P.~69 in
{\em Bell's Theorem, Quantum Theory, and Conceptions of the Universe},
M.~Kafatos, ed., Kluwer, Dordrecht, The Netherlands (1989).
Reproduced as arXiv:0712.0921 (2007).

\bibitem{Mermin90}
N.D.~Mermin,
{\em What's wrong with these elements of reality?}
Phys.~Today~{\bf 43}, 9 (June 1990).

\bibitem{Gurvits03}
L.~Gurvits,  {\em Classical deterministic complexity of Edmonds'
problem and quantum entanglement}. In Proceedings of the Thirty-Fifth ACM
Symposium on Theory of Computing (ACM, New York, 2003),
pp.~10-19.

\bibitem{LeinaasMyrheim06}
J.M.~Leinaas, J.~Myrheim and E.~Ovrum,
{\em Geometrical aspects of entanglement},\\
Phys.~Rev.~A {\bf74}, 012313 (2006).

\bibitem{LeinaasMyrheim07}
J.M.~Leinaas, J.~Myrheim and E.~Ovrum,
{\em Extreme points of the set of density matrices with positive
  partial transpose},
Phys.~Rev.~A {\bf76}, 034304 (2007).

\bibitem{Sollid10a}
J.M.~Leinaas, J.~Myrheim and P.{\O}.~Sollid,
{\em Numerical studies of entangled PPT states in composite 
quantum systems},
Phys.~Rev.~A~{\bf81}, 0062329 (2010).

\bibitem{Sollid10b}
J.M.~Leinaas, J.~Myrheim and P.{\O}.~Sollid,
{\em Low-rank extremal positive-partial-transpose states and
unextendible product bases},
Phys.~Rev.~A~{\bf81}, 0062330 (2010).

\bibitem{Horodecki96} M.~Horodecki, P.~Horodecki and R.~Horodecki,
{\em Separability of mixed states: necessary and sufficient conditions},
Phys.~Lett.~A {\bf 223}, 1 (1996).

\bibitem{Peres96} A.~Peres,
{\em Separability Criterion for Density Matrices},
Phys.~Rev.~Lett.~{\bf 77}, 1413 (1996).

\bibitem{Horodecki00}
P.~Horodecki, M.~Lewenstein, G.~Vidal, and I.~Cirac, 
{\em Operational criterion and constructive checks for the
  separability of low-rank density matrices},
Phys.~Rev.~A, {\bf 62}, 032310 (2000).

\bibitem{Bennett99}
C.H.~Bennett, D.P.~DiVincenzo, T.~Mor, P.W.~Shor, J.A.~Smolin, and
B.M.~Terhal,
{\em Unextendible Product Bases and Bound Entanglement},
Phys.~Rev.~Lett.~{\bf 82}, 5385 (1999).

\bibitem{DiVincenzo03}
 D.P.~DiVincenzo, T.~Mor, P.W.~Shor, J.A.~Smolin, and B.M.~Terhal, 
{\em Unextendible Product Bases, Uncompletable Product Bases and
  Bound Entanglement},\\
Commun.~Math.~Phys.~{\bf 238}, 379 (2003).

\bibitem{SollidLeinaas10}
P.{\O}.~Sollid, J.M.~Leinaas, and J.~Myrheim,
{\em Unextendible product bases and extremal density matrices with positive
partial transpose}.  arXiv:1104.1318.


\end{thebibliography}
\end{document}